# Bridging ambient- and high-pressure superconductivity in $La_2LnNi_2O_7$ films

Motoki Osada[1,2,*], Chieko Terakura[3], Shusaku Imajo[4,5], Jean-Baptiste Morée[3], Akiko Kikkawa[3], Masamichi Nakajima[3], Hsiao-Yi Chen[6], Yusuke Nomura[6,7], Koichi Kindo[4], Ryotaro Arita[3,8], Yoshinori Tokura[2,3,9], Atsushi Tsukazaki[1,2,6,*]

*[1] Quantum-Phase Electronics Center (QPEC), The University of Tokyo, Hongo, Tokyo, Japan*

*[2] Department of Applied Physics, The University of Tokyo, Hongo, Tokyo, Japan*

*[3] RIKEN Center for Emergent Matter Science (CEMS), Wako, Saitama, Japan*

*[4] Institute for Solid State Physics, The University of Tokyo, Kashiwa, Chiba, Japan*

*[5] Department of Advanced Materials Science, The University of Tokyo, Kashiwa, Chiba, Japan*

*[6] Institute for Materials Research (IMR), Tohoku University, Sendai, Miyagi, Japan*

*[7] Advanced Institute for Materials Research (WPI-AIMR), Tohoku University, Sendai, Miyagi, Japan*

*[8] Department of Physics, The University of Tokyo, Hongo, Tokyo, Japan*

*[9] Tokyo College, The University of Tokyo, Hongo, Tokyo, Japan*

## Abstract

The discovery of high critical-temperature $T_c$ superconductivity near 80 K in bilayer nickelates under high pressure has sparked extensive studies. While superconductivity exceeding 40 K was subsequently discovered at ambient pressure in compressively-strained films, the relationship between ambient- and high-pressure regimes remains an open question. Here we present a systematic investigation of superconductivity in compressively-strained $La_2LnNi_2O_7$ films ($Ln$ = lanthanides) at ambient and high pressures. The normal-state resistivity at ambient pressure, revealed by suppressing superconductivity with magnetic fields of 59 T, tends toward $T^2$ behaviour. Under high pressure in a cubic-anvil cell, $T_c$ was enhanced from 41–42 K at ambient pressure to 67–73 K at 16 GPa. On the other hand, lattice compression induced by $Ln$ substitution, which may mimic effects of pressure, lowers $T_c$. In both cases, $T_c$ correlates with the evolution of normal-state transport between $T^2$ and $T$-linear behaviour, offering insight into the interplay between lattice structure and superconductivity in bilayer nickelates.

---

osada@ap.t.u-tokyo.ac.jp, tsukazaki@ap.t.u-tokyo.ac.jp

## Main text

A signature of high-$T_c$ superconductivity near 80 K was observed in the Ruddlesden-Popper phase compound $La_3Ni_2O_7$ under pressure above 14 GPa (Ref.1). Subsequent investigations confirmed zero resistance and diamagnetic signals in single crystals and polycrystals of $La_2LnNi_2O_7$ (*Ln* = lanthanides, e.g., Pr)[2,3,4,5,6,7,8], and as well as in trilayer nickelates[9,10]. The bilayer nickelate is marked by two $NiO_6$ octahedral layers separated by lanthanide rock-salt layers depicted in Fig. 1a, which hosts an average Ni 3*d* electron count of approximately 7.5, possibly forming a half-filled correlated $d_{3z^2-r^2}$ band and a quarter-filled $d_{x^2-y^2}$ band[11,12,13]. Consequently, the significance of interlayer superexchange, enabled by substantial out-of-plane hopping, $t_\perp$, through apical oxygen $2p_z$ orbitals linking the two adjacent $NiO_2$ planes, has been theoretically proposed[11,12,13]. This mechanism is distinct from the single-band $3d^9$ configuration with a half-filled $d_{x^2-y^2}$ band in high-$T_c$ copper oxides[14,15] and $3d^6$ configuration in Fe-based superconductors[16], offering fresh insight into the pairing interactions in high-$T_c$ superconductors. Nevertheless, the high pressures required to stabilize the superconducting state have limited the use of key experimental techniques, such as photoemission spectroscopy and tunneling spectroscopy measurements, leaving the electronic structure and pairing mechanism largely unexplored. Compressively strained bilayer nickelate thin films have exhibited superconductivity at ambient pressure, with an onset $T_c$ above 40 K (Refs.17,18), enabling spectroscopic studies of the superconducting state[19,20,21,22]. The Fermi surface has been experimentally revealed with multi-orbital characteristics[21,22]. Although the electronic state at ambient pressure has been studied, no experiment has examined its evolution under pressure. Whether the superconducting state in strained films is intrinsically consistent with that observed in bulk under high pressure remains an open question.

Bulk studies have been basically conducted under hydrostatic pressure (top panel of Fig. 1b)[2,4,5], where both the *a*-/*b*-axis and *c*-axis lattice constants decrease under isotropic pressure. In contrast, the film studies by epitaxial strain imposes in-plane biaxial compression along the *a*,*b* plane, accompanied by elongation along the *c*-axis owing to the Poisson effect (bottom panel of Fig. 1b). Hereafter, for comparison between bulk and thin films, we use the pseudo-tetragonal in-plane lattice constant ($\sqrt{a^2+b^2}/2$) as the *a*-axis lattice constant. The relation between *c*- and *a*-axis lattice constants of bulk and film are summarized in Fig. 1c. The ratio of *c*- and *a*-axis lattice constant (*c*/*a*) of bulk at ambient pressure is approximately 5.3 with $a$ = 3.83 Å and $c$ = 20.5 Å. By applying a pressure of 20 GPa, superconductivity around 80 K is observed in $La_2PrNi_2O_7$ (Ref. 4), where the lattice is squeezed to $a$ = 3.70 Å and $c$ = 19.5 Å while nearly maintaining $c/a$ ~ 5.3. Superconductivity tends to appear once the $a$-axis falls below ~3.76 Å, and the observed increase in $T_c$ under hydrostatic pressure is consistent with a positive correlation between further lattice compression and $T_c$. The in-plane lattice of compressively strained films is close to 3.75 Å that is governed by $a$-axis lattice constant of $SrLaAlO_4$ substrate. As the $c$-axis length increases to approximately 20.5 Å, the $c/a$ ratio approaches 5.5. Based on the lattice constants of the film at ambient condition, the pressure effect to compress the lattice is anticipated with the assumption of comparable Young's modulus (broken arrow in Fig. 1c). Both axis lengths are compressed approximately 4−5%, a value not achievable by strain alone in thin films. In this study, we employ a unified approach combining state-of-the-art thin-film fabrication and high-pressure measurements to investigate the evolution of superconductivity under lattice compression in compressively strained films. We also present a systematic lanthanide-substitution series, stabilized by epitaxial strain, and show that the ambient-pressure $T_c$ evolves systematically in correlation with the transport coefficients.

## Structural characterization

We prepared $La_2LnNi_2O_7$ films ($Ln$: Y, La, Pr, Nd, Sm, Eu, Dy) on $SrLaAlO_4$ substrates by pulsed laser deposition and applied ex-situ ozone annealing to gain experimental insight into the stabilization of superconductivity in bilayer nickelates (see Extended Data Figs. 1−2 for the estimation of $c$-axis and $a$-axis lengths of the films of a lanthanide series and for the detection of four-fold symmetry indicating tetragonal structure for $La_2SmNi_2O_7/SrLaAlO_4$)[23]. Among the lanthanide series, $La_2SmNi_2O_7$ thin films were the most successfully stabilized under our fabrication conditions, so we focus on this compound, while noting that stronger 004/008 reflections remain a desirable indicator of structural quality and that further optimization is warranted. Figure 1d shows a cross-sectional high-angle annular dark-field (HAADF) scanning transmission electron microscopy (STEM) image of a $La_2SmNi_2O_7$ film, revealing a highly coherent bilayer 2222-structure with minimal disorder over a wide lateral range, with a thickness of ~4 nm that maintains coherent epitaxial strain from the $SrLaAlO_4$ substrate. Energy-dispersive X-ray spectroscopy (EDS) detects the elemental distribution with sub-atomic resolution (Fig. 1e). Based on the atomic contrast in STEM-EDS mapping and profiles (Extended Data Fig. 2c–e), La and Sm atoms are preferentially located within the $NiO_6$ bilayers and in the rock-salt layers (*Ln* (1) and *Ln* (2) in Fig. 1a), respectively. Note that the STEM-EDS color maps are qualitative and reflect relative enrichment, not absolute atomic fractions. While Sm is preferentially located on the *Ln* (2) sites, the nominal composition remains La:Sm = 2:1. A similar Sm distribution was reported in bulk $La_2SmNi_2O_7$ (Ref. 24), indicating stabilization via lanthanide substitution, as in $(La,Pr)_3Ni_2O_7$ (Refs. 4, 18, 19).

## Ambient-pressure superconductivity

The ozone-annealed $La_2SmNi_2O_7$ thin film on $SrLaAlO_4$ (001) substrate exhibits a superconducting transition with an onset temperature $T_c^{onset}$ = 41 K, which is comparable to values reported in previous studies on $La_3Ni_2O_7$ and $(La,Pr)_3Ni_2O_7$ thin films[17,18]. The resistance drop was also comparably sharp, with the resistivity decreasing to 10% of its initial value at 28 K and reaching zero resistance at 6 K (Fig. 2a). Without ozone-annealing, the insulating behaviour was observed in the film (the inset of Fig. 2a), which is also consistent to that in the previous studies[17,18]. Figure 2b displays the temperature dependence of resistivity $\rho(T)$ curves for the $La_2LnNi_2O_7$ thin films. Superconducting transitions appeared for *Ln* = Pr, Nd, Sm, Eu, and Dy, with onset temperatures $T_c^{onset}$ of 47 K, 42 K, 41 K, 10 K, and 20 K, respectively, which was confirmed by the suppression of the resistive downturn under applying out-of-plane magnetic field (Extended Data Fig. 3). In contrast, for *Ln* = La and Y, resistance upturn was observed, which may be related to extrinsic disorders such as insufficient stoichiometry control of La/Ni ratio for La [Ref.18] and/or small lattice size for Y. The $T_c^{onset}$ for the films is plotted in Fig. 2c as a function of the *c*-axis lattice constant with data from early studies[17,18,19]. Overall, our results suggest a correlation between the *c*-axis lattice constant and $T_c^{onset}$ across the *Ln* series, although high crystalline quality and proper oxygen stoichiometry are also essential[25]. In addition, as shown in the inset of Fig. 2c, Hall coefficients, $R_H$, are systematically correlated with *c*-axis constant presenting sign reversal at 20.7 Å corresponding to *Ln* = Pr (also see Extended Data Fig. 4 for temperature dependence of $R_H$). In Fig. 2d, the relation between $T_c^{onset}$ and $R_H$ reveal that a superconducting phase emerges at $R_H > -0.39 \times 10^{-3}$ $cm^3C^{-1}$ across sign reversal of $R_H$ from negative to positive. Considering the incipient band model based on hybridized $d_{3z^2-r^2}$ bands close to Fermi energy[13], the clear trend of high $T_c$ at small negative $R_H$ or positive $R_H$ likely reflects the large contribution of hole-like $d_{3z^2-r^2}$ band closing to $E_F$. These trends are more naturally attributed to changes in the multicarrier Fermi surface associated with lanthanide substitution, rather than to dependence on the $c$-axis parameter alone. In a simple multiband picture with $d_{3z^2-r^2}$ and $d_{x^2-y^2}$ states, lanthanide substitution redistributes carriers between bands, correlating with the resistivity and the evolution of $T_c$.

To further probe the normal-state transport hidden by the superconducting state, we perform electrical transport measurements in the $La_2SmNi_2O_7$ film under high magnetic fields. Figure 2e shows the $\rho(T)$ curves measured under magnetic fields of up to 59 T applied along the $c$-axis (out-of-plane direction; data for in-plane and out-of-plane magnetic fields are provided in Extended Data Fig. 5). The superconducting state is progressively suppressed with increasing field and is destroyed at 59 T, where the low-temperature normal-state transport in the range of 7–40 K follows a power-law behaviour, $\rho = \rho_o + AT^{\alpha}$, with an exponent $\alpha \sim 1.86$ ($\alpha \sim 1.70$ when fitted over for 70–120 K). The fitting range of 7–40 K was chosen to characterize the low-temperature field-induced normal state while reducing higher-temperature contributions from additional scattering channels, such as electron–phonon scattering. This value deviates from $T$-linear behaviour, suggesting possibly crossover regime rather than a simple Fermi-liquid regime. For comparison, we also analyze the data using a $T^2$ dependence. The coefficient $A$ in the fitting curve $\rho = \rho_0 + AT^2$ is 5.4 $n\Omega cmK^{-2}$ in the 7–40 K range and 4.0 $n\Omega cmK^{-2}$ in the 50–90 K range. The $AT^2$ term reflects electron-electron scattering and is proportional to the square of electronic specific coefficient depending on the Kadowaki-Woods ratio. The values obtained in this study are comparable to those reported for non-superconducting overdoped $La_{1.7}Sr_{0.3}CuO_4$ (2.5–4.0 $n\Omega cmK^{-2}$)[26], implying the weakly-correlated metallic transport. Considering the itinerant nature of the transport behaviour, the superconductivity in these ambient-pressure films may be interpreted within a framework of superconducting mechanisms mediated by bosonic fluctuations, as has been discussed for cuprate[27] and iron-based superconductors[28]. Additional analysis of the normal-state transport reveals that $Ln$-substituted samples with higher $T_c$ tend to exhibit normal-state resistivity exponents $\alpha$ approaching unity (Extended Data Fig. 4). While other scattering processes can mimic $T^{\alpha}$ behaviour, this observation suggests an empirical correlation between enhanced superconductivity and more nearly $T$-linear resistivity. We note, however, that this trend should be interpreted with caution, because the $Ln$-substituted samples may also differ in disorder levels, oxygen stoichiometry/affinity, and carrier concentration, all of which can affect both $\alpha$ and $T_c$. Thus, the $\alpha$–$T_c$ correlation may partly reflect differences in sample quality rather than solely indicating a fundamental electronic crossover. Nevertheless, the observed trend remains consistent with the possibility that electronic correlations tuned by structural or chemical modifications play important roles in stabilizing high-$T_c$ superconductivity in bilayer nickelates.

Figure 2f and g show the in-plane upper critical field $H_{c2||ab}$ and out-of-plane upper critical field $H_{c2||c}$ with color map of resistivity, respectively. Here, we define $H_{c2||ab(c)}$ as the magnetic field with 90% and 50% of the normal resistance. The dashed curves represent fits using the Abrikosov-Gorkov model[29]. Extrapolating $T \rightarrow 0$ K, the values of $\mu_0 H_{c2||ab}$ and the $\mu_0 H_{c2||c}$, are estimated to be 86 T and 53 T, respectively, where $\mu_0$ is vacuum permittivity. The in-plane and out-of-plane coherence lengths are $\xi_{||}$ = 25 Å and $\xi_{\perp}$ = 15 Å, respectively. Given that the $c$-axis constant is 20.5 Å, the shorter out-of-plane coherence length indicates a quasi-two-dimensional superconductivity in the $La_2SmNi_2O_7$ film. The superconducting anisotropy, $\gamma_H = H_{c2||ab}/H_{c2||c} = 1.62$, and the ratios $\mu_0 H_{c2||ab}/T_c = 2.26$ T/K and $\mu_0 H_{c2||c}/T_c = 1.39$ T/K, which are larger and smaller than the Pauli limit of 1.86 T/K, respectively[30], further support the anisotropic character.

## High-pressure superconductivity

We applied hydrostatic pressures up to 20 GPa using a cubic-anvil cell (see Extended Data Fig. 6 for sample geometry for high-pressure measurement) with measuring the electrical resistivity of the $La_2SmNi_2O_7$ and $La_2NdNi_2O_7$ films, which exhibit $T_c$ = 41 K and 42 K at ambient pressure, respectively. As shown in Fig. 3a, $\rho(T)$ decreases monotonically with pressure, reflecting enhanced orbital hybridization due to lattice squeezing. The samples were measured without severe damage under systematic increase of pressure (Extended Data Fig. 6b). The intermediate $\alpha$ in $\rho$–$T$ behaviour of the normal-state resistivity approaches unity with increasing pressure to 16 GPa (Extended Data Fig. 7), which may reflect the $T$-linear behaviour accompanying superconductivity as reported in previous studies on bulk[1,2,3] and thin films[31]. As shown in the magnified transition region in Fig. 3b (offset added for clarity), two trends emerge with increasing pressure. For Sm-substituted film $La_2SmNi_2O_7$, the $T_c$ initially rises to 45 K at 4 GPa. Above 8 GPa, in addition to the systematic increase trend, other transitions newly appear, defined as $T_c^{high}$ and $T_c^{low}$, respectively. This two-step transition is clearly observed in the temperature-dependent derivative of resistivity, as shown in Figs. 3c and d and Extended Data Fig. 6. At each transition, the resistivity drops as much as 50%, finally reaching a value close to zero. The $T_c^{high}$ increases with pressure, reaching 67 K at 16 GPa, while $T_c^{low}$ decreases to 28 K at 20 GPa. To verify this pressure evolution of $T_c$, we performed high-pressure measurements on the Nd-substituted film $La_2NdNi_2O_7$. As shown in Fig. 3e, $\rho(T)$ decreases with increasing pressure, with $T_c$ reaching 73 K at 16 GPa. The Nd-substituted film also exhibits a two-step transition above 12 GPa (Fig. 3f–h).

The $T_c^{onset}$ is plotted as a function of pressure in Fig. 4a with those of bulk values in previous studies[1,4] (see Extended Data Fig. 6 for a color map of normalized $\rho(T)$ at 100 K). Superconductivity at ambient pressure persists below 8 GPa, implying that the lattice squeezing enhances Cooper pair formation with increasing conductivity. As pressure increases, $T_c^{high}$ approaches the high-$T_c$ observed in bulk. Importantly, superconductivity observed in bulk above 8 GPa (Refs. 1, 4) coincides with the onset $T_c^{high}$ in the film, suggesting that $c$-axis compression may be one of the factors contributing to the difference in $T_c$ between ambient-pressure films and high-pressure bulk samples. The $T_c^{high/low}$ at 20 GPa are plotted as a function of the ambient-pressure $c/a$ lattice constant ratio at ambient pressure in Fig. 4b, along with data for $La_3Ni_2O_7$ films from a previous study[31]. This indicates that both $La_2SmNi_2O_7$ and $La_2NdNi_2O_7$ films follow the same trend, in which a larger $c/a$ ratio at ambient pressure correlates with a higher $T_c$ under pressure. A consistent tendency is also observed in the relation of onset $T_c$ at ambient pressure, where an elongated $c$-axis at a fixed in-plane lattice constant is associated with enhanced $T_c$ (Fig. 2c). Importantly, the observed correlation between $c/a$ and $T_c$ at $P = 20$ GPa in the *Ln*-substituted thin-film series should be understood within the epitaxial strain framework of films, in which the in-plane lattice constant is constrained by the substrate while the out-of-plane lattice parameter is relaxed. In this context, for strained films, the initial lattice configuration prior to pressure application, characterized by a relatively large $c/a$ ratio, may provide a favorable structural environment for superconductivity. Due to the technical limitations of our cubic-anvil cell, advanced measurements such as magnetic-field-dependent transport and in situ structural probes cannot be applied under high pressure. Future diamond-anvil cell studies combined with magnetic-field or spectroscopic techniques will therefore be essential to further verify and elucidate the evolution of superconductivity, as well as the corresponding electronic and structural states, across the transition from ambient to high pressure.

## Discussion

Here, we discuss the similarities and differences between superconductivity under *Ln* substitution at ambient pressure and superconductivity under hydrostatic pressure in bilayer nickelate thin films, based on experimental observations. In the former case, lattice compression involves only a change in the $c$-axis length, whereas the latter involves compression along both the $a$ and $c$ axes. Their effects on superconductivity are found to be distinct. For *Ln* substitution, the parameter $\alpha$ of $T^{\alpha}$ describing the temperature dependence of electrical resistivity increases with $c$-axis compression and $T_c$ decreases (Extended Data Fig. 4). In contrast, under hydrostatic pressure, the opposite trend is observed: lattice compression tends to reduce $\alpha$ and increase $T_c$ (Extended Data Fig. 7). Despite these differences, when considering only the relationship between $\alpha$ and $T_c$, smaller $\alpha$ tends to correspond to higher $T_c$ in both cases. We note that the extracted value of $\alpha$ can be affected by the choice of fitting temperature range; nevertheless, fits performed using a uniform 70–120 K window indicate that the qualitative trend is largely preserved. This interesting similarity indicates that deviations from Fermi-liquid-like behaviour are associated with $T_c$ enhancement in both cases.

This systematic investigation reveals a complex interplay among the lattice parameters $a$, $c$, and $c/a$ in the $T_c$ landscape. To examine how lattice constants affect the electronic structure, density functional theory (DFT) calculations were performed over a relatively wide range of $a$ and $c$ values (Extended Data Fig. 8). While $c/a$ shows a positive correlation with changes in the energy difference $\Delta E$ between the $d_{x^2-y^2}$ and $d_{3z^2-r^2}$ orbitals (Extended Data Fig. 8c), the $a$- and $c$-axis lengths show a negative correlation with in-plane hopping and interlayer hopping, respectively (Extended Data Fig. 9). Changes in these microscopic parameters can also modify the Fermi surface. In the region corresponding to thin films at ambient pressure, the Fermi surface consists of the $\alpha$ and $\beta$ sheets formed by hybridization of the $d_{x^2-y^2}$ and $d_{3z^2-r^2}$ orbitals (Extended Data Fig. 8a). In contrast, in the region corresponding to bulk samples under hydrostatic pressure, the $\gamma$ sheet originating from the $d_{3z^2-r^2}$ orbital appears (Extended Data Fig. 8b). In this way, changes in the lattice constants lead to complex modifications of the electronic structure in this multi-orbital system. Within the lattice-parameter space expanded by this study, disentangling the roles of $a$, $c$, and $c/a$ in superconductivity will be an important direction for future research.

Finally, we propose the origins for the observation of two-step transition. A plausible extrinsic origin of the two-step transition is spatial inhomogeneities, notably oxygen deficiency and La/Sm distribution. Although the ozone-annealed films exhibit a dramatic evolution from insulating to metallic conduction (inset of Fig. 2a), spatially inhomogeneous oxygen deficiency within the films cannot be excluded, and potentially pre-existing inhomogeneity in oxygen content may persist. In addition, considering the previous observation of sharp one-step transition in $La_3Ni_2O_7$ film under pressure[31], the La/Sm and La/Nd composition fluctuation also affects the transport property under pressure. Since hydrostatic pressure enhances lattice compression and orbital hybridization, it may amplify such pre-existing inhomogeneity, leading to spatially separated superconducting regions. Another possible extrinsic scenario is non-uniform strain or effective-pressure transmission in the film–substrate system under hydrostatic compression, where the near-interface region and the rest of the ultrathin film may experience different effective strain/pressure conditions. Such non-uniformity could produce spatially distinct superconducting regions with different $T_c$ values, as suggested by related high-pressure studies of broadened superconducting transitions in infinite-layer nickelate thin films[32] and two-step transitions in Bi crystals[33]. While the orbital character of $d_{3z^2-r^2}$ and $d_{x^2-y^2}$ (refs.[34,35,36]) may also become pronounced under pressure and potentially give rise to multiband pairing with different gap scales, direct microscopic evidence is currently unavailable. Further studies on defect chemistry, theoretical modelling, and microscopic measurements will be essential to deepen the physical understanding of these scenarios.

We carried out a unified experimental approach to examine ambient- and high-pressure superconductivity in $La_2\mathit{Ln}Ni_2O_7$ thin films. Our results reveal an interplay between lattice structure and superconductivity. Although high-$T_c$ in thin films can be achieved under high pressure through lattice compression, we note that $c$-axis compression via lanthanide substitution does not produce the same effect at ambient pressure. Despite this contrast, both approaches exhibit a similar tendency in which smaller power-law exponent $\alpha$ of temperature-dependent resistivity accompanies higher $T_c$. Moreover, the anomalous pressure induced two-step superconducting transition may be extrinsic in origin, as discussed above, although intrinsic contributions, such as the evolution of the electronic band structure, remain possible and will require further investigation into orbital contributions, particularly the $d_{x^2-y^2}$ and $d_{3z^2-r^2}$ states. Investigation of distinct strain regimes may further enable structural and electronic factors to be decoupled[37]. The control of structural symmetry and band-filling thorough epitaxial strain and chemical substitution deserves further attention for achieving higher-$T_c$ superconductivity at ambient pressure.

## Acknowledgements

The authors thank M. Kawasaki and M. Nakamura for technical supports and fruitful discussion. STEM observations were made with the cooperation of Y. Kodama, K. Hayasaka, and T. Konno of Analytical Research Core for Advanced Materials, Institute for Materials Research, Tohoku University.

## Funding Statement

A part of this work was supported by Tohoku University in MEXT Advanced Research Infrastructure for Materials and Nanotechnology in Japan (Grant No. JPMXP1224TU0193), Basic Science Research Projects by The Sumitomo Foundation, Toyota Riken Scholar Program by Toyota Physical and Chemical Research Institute, The Kazuchika Okura Memorial Foundation, the RIKEN TRIP initiative (RIKEN Quantum, Advanced General Intelligence for Science Program, Many-body Electron Systems), and JSPS KAKENHI (Grant Nos. JP23K13663, JP24H00190, JP25K00015, JP25H01246, JP25H01250, and JP25H01252). High-field measurements were performed using facilities of the Institute for Solid State Physics, the University of Tokyo (Grant No. 202505-HMBXX-0084). Y.N. acknowledges support from JSPS KAKENHI (Grant Nos. JP23H04869, JP23K03307, and JP25H01506) and MEXT as “Program for Promoting Researches on the Supercomputer Fugaku” (Project ID: JPMXP1020230411).

## Author contributions

M.O. and A.T. conceived the project. M.O. fabricated and characterized nickelate films. S.I. and K.K. performed the high-magnetic field measurement and analysis. M.O., C.T., A.K., M.N., Y.T., and A.T. performed the high-pressure measurement and analysis. J.-B.M., H.-Y.C., Y.N., and R.A. conducted theoretical calculation. M.O. and A.T. wrote the manuscript with input from all the authors.

## Competing interests

The authors declare no competing interests.

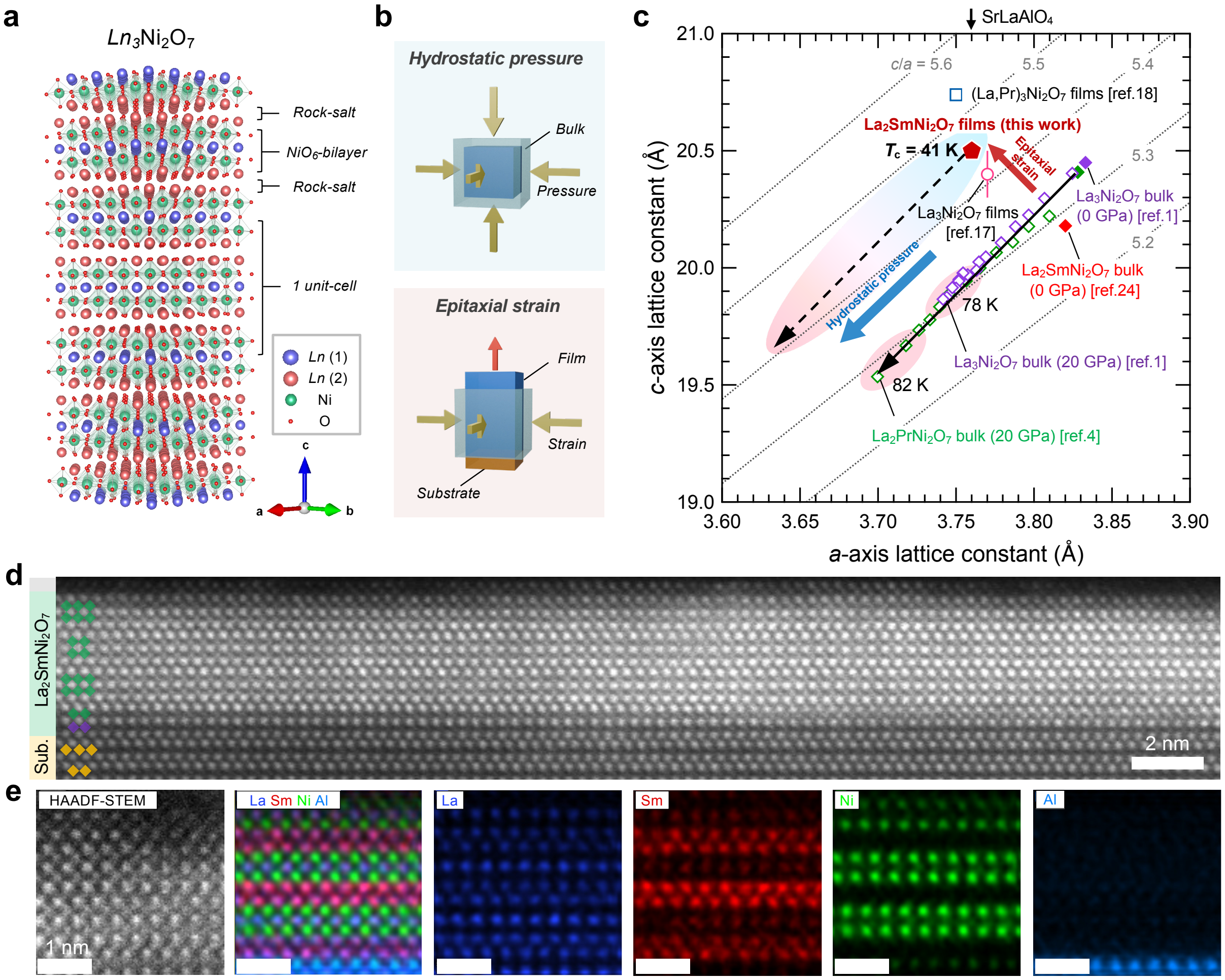

**Fig. 1 | Crystalline lattices of superconducting bilayer nickelates and characterization of $La_2SmNi_2O_7$ thin films. a,** crystal structure of $La_2LnNi_2O_7$. **b,** top and bottom panel shows sample schematics under hydrostatic pressure (top) and epitaxial strain (bottom). **c,** the relation between *c*-axis and *a*-axis lattice constants for bulk values in previous studies[1,4,24] and the value of thin films[17,18]. Broken lines correspond to the *c*/*a* ratio. Arrows indicate the lattice squeezing under pressure. Note that two schematic trajectories are expected for pressurized films: evolution at nearly constant *c*/*a*, and convergence of *c*/*a* toward ~5.3 due to substrate constraint. **d,** high-resolution HAADF-STEM image of $La_2SmNi_2O_7$ film on $SrLaAlO_4$ substrate taken along $La_2SmNi_2O_7$ [110] axis. **e,** STEM-EDS mapping for La, Sm, Ni, and Al of $La_2SmNi_2O_7$ film on $SrLaAlO_4$ substrate. Note that STEM-EDS maps are qualitative, showing relative enrichment rather than absolute atomic fractions. Sm is preferentially on *Ln* (2), but the nominal composition is La:Sm = 2:1.

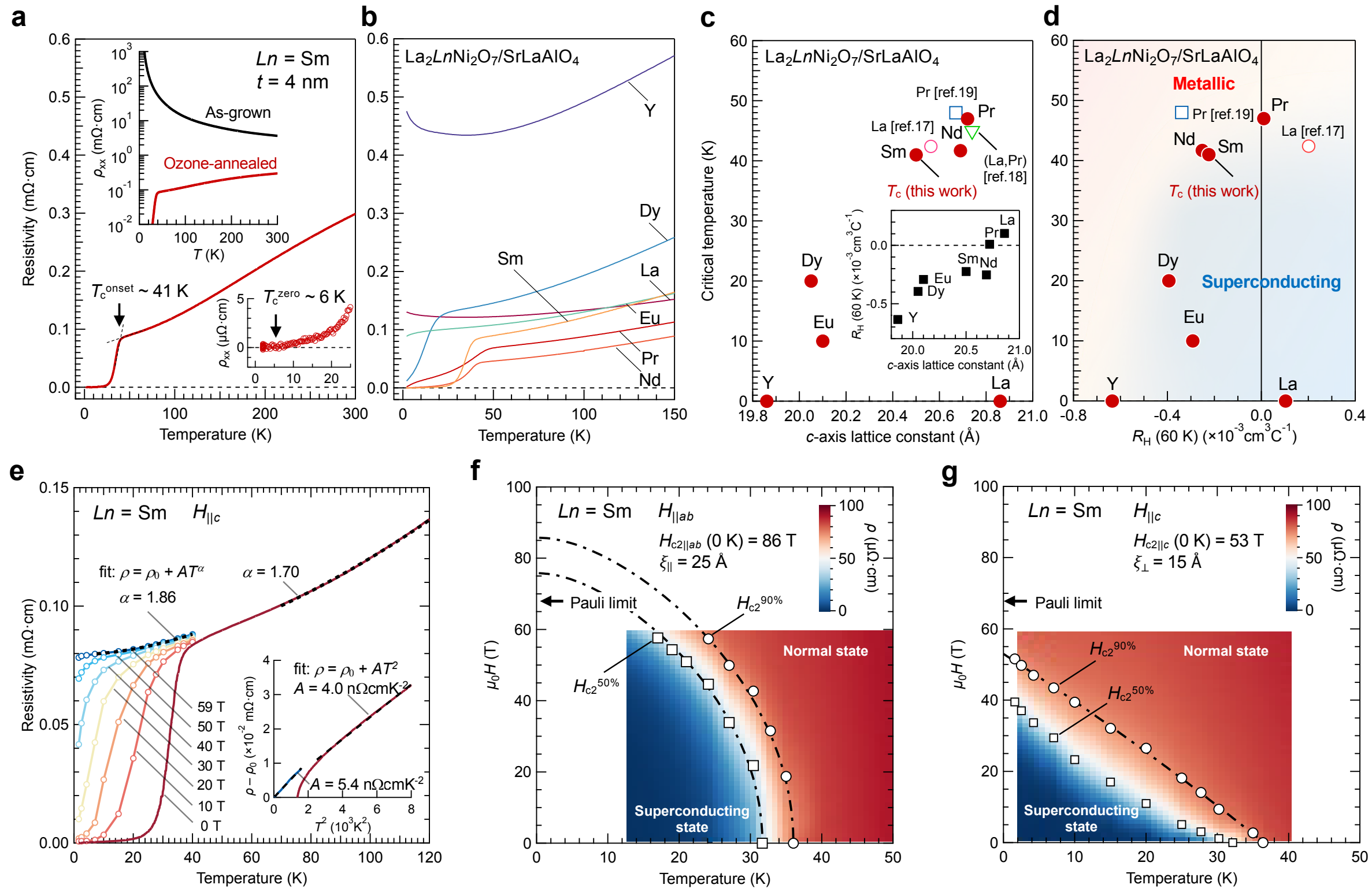

**Fig. 2 | Superconductivity of $La_2LnNi_2O_7$ films at ambient pressure. a,** Temperature dependence of resistivity of $La_2SmNi_2O_7$ film. The top inset shows the logarithmic scale of resistivity of the film before and after ozone annealing. The bottom inset shows a low-temperature resistivity of the film indicating zero resistance at approximately 6 K. **b,** Temperature dependence of resistivity for the $La_2LnNi_2O_7$ films (*Ln* = La, Pr, Nd, Sm, Eu, Dy, Y). **c,** Superconducting transition temperature as a function of *c*-axis lattice constant. Closed red circles are the data in this study and open symbols are the data in previous studies[17,18,19]. The inset shows the Hall coefficient at 60 K, $R_H$ (60 K), as a function of *c*-axis lattice constant. **d,** Superconducting transition temperature as a function of Hall coefficient at 60 K. Closed red circles are the data in this study and open symbols are the data in previous studies[17,19]. **e,** Temperature dependence resistivity of $La_2SmNi_2O_7$ film under magnetic field up to 59 T. The inset shows the normalized resistivity as a function of $T^2$. Broken curves in main panel represent fitting line of $\rho = \rho_o + AT^{\alpha}$. Broken line in the inset shows $\rho = \rho_o + AT^2$. **f** and **g,** Counter plot of resistivity as functions of temperature and magnetic field along *ab*-plane and *c*-axis direction, respectively. Open circle and square correspond to the 90 % and 50 % value of normal resistance, respectively. The linear temperature dependence of $H_{c2||c}$ indicates that orbital depairing dominates in a field applied perpendicular to the plane, while the $(T_c\text{–}T)^{1/2}$ dependence of $H_{c2||ab}$ near $T_c$ suggests that the superconductivity is mainly destabilized by the paramagnetic pair breaking effect.

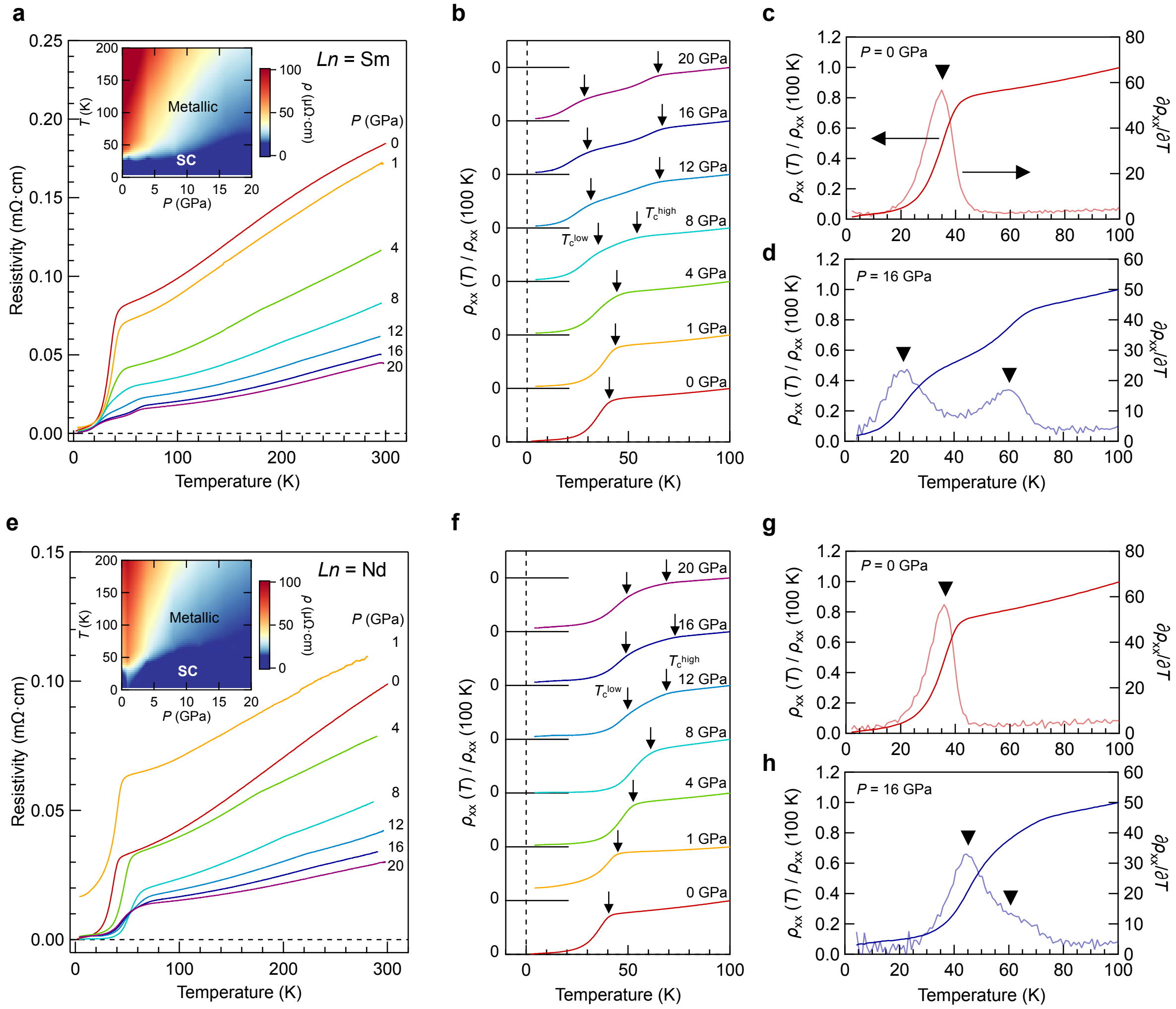

**Fig. 3 | Superconductivity of $La_2LnNi_2O_7$ film under high pressure. a**, Temperature dependence of resistivity under high pressure up to 20 GPa for *Ln* = Sm. The inset is a color map of resistivity as a function of temperature and pressure. **b**, The magnified resistivity data below 100 K for *Ln* = Sm. The arrows correspond to the transition temperature. **c** and **d**, The normalized resistivity at 100 K (left axis) and the temperature-dependent deviation of resistivity (right axis) of $La_2SmNi_2O_7$ films under 0 and 16 GPa, respectively. Triangles indicate resistive transitions. **e**, Temperature dependence of resistivity under high pressure up to 20 GPa for *Ln* = Nd. **f**, The magnified resistivity data below 100 K for *Ln* = Nd. **g** and **h**, The normalized resistivity at 100 K (left axis) and the temperature-dependent deviation of resistivity (right axis) of $La_2NdNi_2O_7$ films under 0 and 16 GPa, respectively. Triangles indicate resistive transitions.

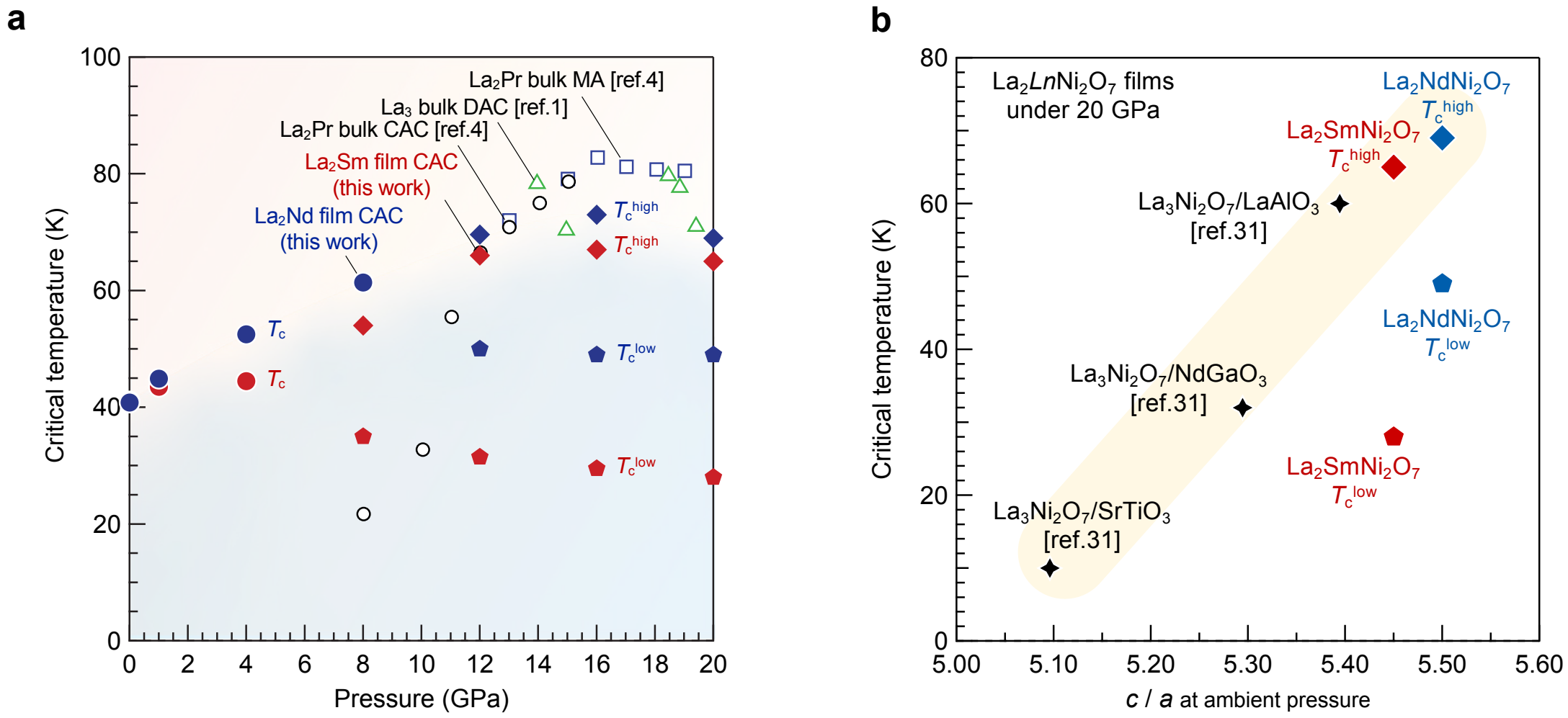

**Fig. 4 | Pressure phase diagram of superconducting $La_2$*Ln*$Ni_2O_7$ films. a,** Transition temperature of $La_2SmNi_2O_7$ and $La_2NdNi_2O_7$ films as a function of pressure. Closed symbols are the data in this study and open symbols are that in previous studies[1,4]. DAC, CAC, and MA denote a diamond-anvil cell, cubic-anvil cell, and multi-anvil, respectively. **b,** The critical temperatures of $T_c^{high}$ (closed diamonds) and $T_c^{low}$ (closed pentagons) for *Ln* = Sm (red markers) and Nd (blue markers) under 20 GPa as a function of ambient-pressure *c*/*a* ratios. Diamonds (black markers) are the data of $La_3Ni_2O_7$ films grown on $SrTiO_3$, $NdGaO_3$, and $LaAlO_3$ substrates under 20 GPa (Ref.31).

## References

1. Sun, H. *et al.* Signatures of superconductivity near 80K in a nickelate under high pressure. *Nature* **621**, 493–498 (2023).
2. Hou, J. *et al.* Emergence of high-temperature superconducting phase in pressurized $La_3Ni_2O_7$ crystals. *Chin. Phys. Lett.* **40**, 117302 (2023).
3. Zhang, Y. *et al*. High-temperature superconductivity with zero resistance and strange-metal behaviour in $La_3Ni_2O_{7-\delta}$. *Nat. Phys.* **20**, 1269–1273 (2024).
4. Wang, N. *et al*. Bulk high-temperature superconductivity in pressurized tetragonal $La_2PrNi_2O_7$. *Nature* **634**, 579–584 (2024).
5. Wang, G. *et al.* Pressure-induced superconductivity in polycrystalline $La_3Ni_2O_{7-\delta}$. *Phys. Rev. X.* **14**, 011040 (2024).
6. Dong, Z. *et al*. Visualization of oxygen vacancies and self-doped ligand holes in $La_3Ni_2O_{7-\delta}$. *Nature* **634**, 579–584 (2024).
7. Yang, J. *et al*. Orbital-dependent electron correlation in double-layer $La_3Ni_2O_{7-\delta}$. *Nat. Commun.* **15**, 4373 (2024).
8. Chen, X. *et al.* Electronic and magnetic excitations in $La_3Ni_2O_7$. *Nat. Commun.* **15**, 9597 (2024).
9. Sakakibara, H. *et al.* Theoretical analysis on the possibility of superconductivity in the trilayer Ruddlesden-Popper nickelate $La_4Ni_3O_{10}$ under pressure and its experimental examination: Comparison with $La_3Ni_2O_7$. *Phys. Rev. B* **109**, 144511 (2024).
10. Zhu, Y. *et al.* Superconductivity in pressurized trilayer $La_4Ni_3O_{10-\delta}$ single crystals. *Nature* **631**, 531–536, (2024).
11. Nakata, M., Ogura, D., Usui, H. & Kuroki, K. Finite-energy spin fluctuations as a pairing glue in systems with coexisting electron and hole bands. *Phys. Rev. B* **95**, 214509 (2017).
12. Maier, T. A., Mishra, V., Balduzzi, G. & Scalapino, D. J. Effective pairing interaction in a system with an incipient band. *Phys. Rev. B* **99**, 140504(R) (2019).
13. Sakakibara, H., Kitamine, N. Ochi, M. & Kuroki, K. Possible High $T_c$ Superconductivity in $La_3Ni_2O_7$ under High Pressure through Manifestation of a Nearly Half-Filled Bilayer Hubbard Model. *Phys. Rev. Lett.* **132**, 106002 (2024).
14. Bednorz, B. G. & Müller, K. A. Possible high$T_c$ superconductivity in the Ba−La−Cu−O system. *Z. Physik B–Condensed Matter* **64**, 189–193 (1986).
15. Keimer, B., Kivelson, S. A., Norman, M. R., Uchida, S. & Zaanen, J. From Quantum Matter to High-Temperature Superconductivity in Copper Oxides. *Nature* **518**, 179–186 (2015).
16. Fernandes, R. M. *et al.* Iron Pnictides and Chalcogenides: a New Paradigm for Superconductivity. *Nature* **601**, 35–44 (2022).
17. Ko, E. K. *et al.* Signatures of ambient pressure superconductivity in thin film $La_3Ni_2O_7$. *Nature* **638**, 935–940 (2025).
18. Zhou, G. *et al*. Ambient-pressure superconductivity onset above 40 K in $(La,Pr)_3Ni_2O_7$ films. *Nature* **640**, 641–646 (2025).
19. Yidi, L. *et al.* Superconductivity and normal-state transport in compressively strained $La_2PrNi_2O_7$ thin films. *Nat. Mater.* **24**, 1221–1227 (2025).
20. Bhatt, L. *et al.* Structural modifications in strain-engineered bilayer nickelate thin films. *Nature* **653**, 76–82 (2026).

21. Li, P. *et al.* Angle-resolved photoemission spectroscopy of superconducting $(La,Pr)_3Ni_2O_7/SrLaAlO_4$ heterostructures. *National Science Review* **12**, nwaf205 (2025).
22. Wang, B. Y. *et al.* Electronic structure of compressively strained thin film $La_2PrNi_2O_7$. Preprint at arxiv.org/abs/2504.16372 (2025).
23. Shannon, R. D. Revised Effective Ionic Radii and Systematic Studies of Interatomic Distances in Halides and Chalcogenides. *Acta Cryst.* A **32**, 751–767 (1976).
24. Li, F. *et al.* Bulk superconductivity up to 96 K in pressurized nickelate single crystals. Nature **649**, 871–878 (2026).
25. Zhuo, G. *et al.* Superconductivity onset above 60 K in ambient-pressure nickelate films. *National Science Review* **13**, nwag151 (2026).
26. Hwang, H. Y. *et al.* Scaling of the temperature dependent Hall effect in $La_{2-x}Sr_xCuO_4$. *Phys.Rev. Lett.* **72**, 2636 (1994).
27. Michon, B. *et al.* Thermodynamic signatures of quantum criticality in cuprate superconductors. *Nature* **567**, 210-222 (2019).
28. Jiang, X. *et al.* Interplay between superconductivity and the strange-metal state in FeSe. *Nat. Phys.* **19**, 365–371 (2023).
29. Abrikosov, A. A. and Gor'kov, L. P. Spin-orbit interaction and the knight shift in superconductors. *Soviet Physics JETP* **15**, 752–757 (1962).
30. Maki, K. & Tsuneto, T. Pauli paramagnetism and superconducting state. *Prog. Theor. Phys.* **31**, 945–956 (1964).
31. Osada, M. *et al.* Strain-tuning for superconductivity in $La_3Ni_2O_7$ thin films, *Commun. Phys.* **8**, 251 (2025).
32. Wang, N. N. *et al.* Pressure-induced monotonic enhancement of Tc to over 30 K in superconducting $Pr_{0.82}Sr_{0.18}NiO_2$ thin films, *Nat. Commun.* **13**, 4367 (2022).
33. Li, Y., Wang, E., Zhu, X., Wen, H.-H. Pressure-induced superconductivity in Bi single crystals, *Phys. Rev. B* **95**, 024510 (2017).
34. Geisler, B., Hamlin, J. J., Stewart, G. R., Hennig, R. G. & Hirschfeld, P. J. Fermi surface reconstruction and enhanced spin fluctuations in strained $La_3Ni_2O_7$ on $LaAlO_3(001)$ and $SrTiO_3(001)$, *Phys. Rev. B* **112**, L100506 (2025).
35. Oh, H. *et al*. High spin, low spin or gapped spins: magnetism in the bilayer nickelates, *Phys. Rev. B* **113**, 024430 (2026).
36. Fan, S. *et al*. Superconducting gap structure and bosonic mode in $La_2PrNi_2O_7$ thin films at ambient pressure, Preprint at arxiv.org/abs/2506.01788 (2025).
37. Zhao, Y.-F. & A. Botana, A. S. Electronic structure of Ruddlesden-Popper nickelates: strain to mimic the effects pressure, *Phys. Rev. B* **111**, 115154 (2025).

## Methods

**Sample preparation:** $La_2LnNi_2O_7$ films ($Ln$ = La, Pr, Nd, Sm, Eu, Dy, Y) were fabricated on single-crystalline $SrLaAlO_4$ (001) substrates using pulsed-laser deposition. $La_2LnNi_2O_7$ targets were prepared by mixing precursor powders ($Ln_xO_y$ and NiO) and sintering them at 1200°C for 12 hours, repeated twice. $La_2LnNi_2O_7$ polycrystalline targets were ablated by a KrF excimer laser (wavelength 248 nm). Substrates were pre-annealed at 650°C in an oxygen partial pressure of $1 \times 10^{-6}$ Torr to obtain an atomically flat surface. During growth, the substrate temperature was fixed at 650°C and the oxygen partial pressure was 200–300 mTorr. The laser fluence was ~0.7 J/cm$^2$ and a repetition rate was 4 Hz. $SrTiO_3$ capping layer (~1 u.c.) was deposited on the top of films under the same growth condition as that of $La_2LnNi_2O_7$ films. After the deposition, the sample is transferred ex-situ to ozone treatment apparatus (Samco UV-1). Ozone annealing was carried out at 240–300 °C for a total of approximately 100 minutes under a 500 sccm ozone flow (240°C for 30 min, 270°C for 30 min, and 300°C for 40 min). The optimization of this condition was carried out by $La_2SmNi_2O_7$ films. Identical treatment was applied to other $La_2LnNi_2O_7$ films. Based on the Gibbs free energies of the oxidation reactions for the lanthanide and nickel cations, oxidation in these thin films is governed by nickel, rather than by the lanthanide ions. Accordingly, applying an identical ozone treatment to all compositions allows us to evaluate the lanthanide dependence of the electrical conductivity on a common footing. Since all films were treated by ozone-annealing, the difference of $c$-axis lattice constant comes from the chemical pressure effect with lanthanide substitution because of suppression of oxygen deficiency in the film. The films were characterized using x-ray diffraction (XRD) techniques with Cu K$\alpha$ source ($\lambda$ = 1.5406 Å). Crystal structure depicted in Fig. 1a were visualized using the VESTA software[38].

**Scanning transmission electron microscopy:** To prepare cross-sectional lamellas, we employed a focused ion beam (FIB) lift-out procedure. Scanning transmission electron microscopy (STEM) was conducted using a Titan3 60-300 Probe Corrector. Both STEM imaging and Energy Dispersive X-ray Spectroscopy (EDS) mapping were performed at an accelerating voltage of 200 kV.

**Electrical transport measurement:** The measurements of the temperature dependent resistivity $\rho(T)$ were measured using a six-point geometry with Au wire bonded on Au electrodes (30 nm-thick) in physical properties measurement system (PPMS, Quantum Design, Inc.). Au electrode was deposited at room temperature by electron beam evaporation. The Hall effect was measured to be linear up to the highest measured magnetic field of 9 T (–9 T).

**High-magnetic-field electrical resistance measurement:** High-field electrical resistance measurements were carried out using a non-destructive pulse magnet at the International MegaGauss Science Laboratory, Institute for Solid State Physics in the University of Tokyo. The maximum field and pulse duration of the pulse magnet were 60 T and 36 ms, respectively. Electrical resistances were measured by the standard 4 probe method, and a 80 kHz AC current of approximately 10 μA was applied parallel to the conducting plane of thin-film samples. Unless noted otherwise, $H_{c2}$ is defined using the 90% $R_n$ criterion: for each temperature, $H_{c2}$ is the field at which the resistivity reaches $0.9 \times R_n$. Here $R_n$ denotes the normal-state resistivity.

**High-pressure electric resistivity measurement:** The thin-film samples were cut into dimensions of 500–1000 μm in lateral size, and the substrates were mechanically polished to reduce their thickness to 0.3 mm (see Extended Data Fig. 6a for details). Gold wires were bonded to the samples in a van der Pauw geometry using silver paste to ensure reliable electrical contacts. The $\rho(T)$ curves at $P$ = 0 GPa were observed in physical properties measurement system (PPMS) (Quantum Design, Inc.). The measurements of the temperature dependent resistivity $\rho(T)$ under high-pressure was performed under various hydrostatic pressure of 1–20 GPa using a cubic-anvil cell with a liquid pressure-transmitting medium (Daphne oil 7575, Idemitsu Kosan Co., Ltd.). Thin-film samples were mounted in a Taflon capsule and MgO gasket. Press-loading speeds are: 1.6 ml/min in 0–1 GPa, 1.0 ml/min in 1–4 GPa, 1.0 ml/min in 4–8 GPa, 0.8–1.0 ml/min in 0.8–12 GPa, 0.6–0.8 ml/min in 12–16 GPa, and 0.4 ml/min in 16–20 GPa. The identical setup, exhibiting sharp superconducting transition in previous literature, was applied in this study for uniformly pressurizing the film[31]. The temperature dependence of resistance was measured using Keithley 2182A nanovoltmeters and Keithley 6221 source meters in delta mode. Measurements were conducted over a temperature range from 292 K to 4.2 K.

**Theoretical calculations based on density functional theory:**

The flow of calculations is summarized below. We start from given values of the cell parameters $a$ and $c$ of $La_3Ni_2O_7$. First, (i) we use the density functional theory (DFT) framework to perform a structural optimization. This allows to calculate the atomic positions in the unit cell at fixed ($a$, $c$), defined in Extended Data Table 1. Then, we perform (ii) a self-consistent DFT calculation followed by (iii) a non-self-consistent calculation to obtain the structure shown in the Extended Data Fig. 9. The positions of atoms in the unit cell are given in Extended Data Table 1. Then, (iv) we construct the four Wannier orbitals for the low-energy effective Hamiltonian.

There are two orbitals per Ni atom in the unit cell: The $d_{x^2-y^2}$ orbital that results from the hybridization of Ni3$d_{x^2-y^2}$ and in-plane O2$p_\sigma$ atomic orbitals, and the $d_{3z^2-r^2}$ orbital that results from the hybridization of Ni3$d_{3z^2-r^2}$ and apical O2$p_z$ atomic orbitals. We deduce the one-particle part of the low-energy Hamiltonian by calculating the Kohn-Sham Hamiltonian in the basis of Wannier orbitals.

The computational details are the following. On (i, ii, iii), we use the planewave implementation of the DFT in Quantum ESPRESSO (Refs. 39 and 40), the GGA-PBE functional and optimized norm-conserving Vanderbilt pseudopotentials for La, Ni, and O (Refs. 41, 42, 43). We consider a 100 Ry planewave energy cutoff for wavefunctions, and a Fermi-Dirac occupation smearing of 0.002 Ry. We use a $N_k \times N_k \times N_k$ $k$-point grid, with $N_k = 8$ for (i), $N_k = 12$ for (ii), and $N_k = 8$ for (iii). On (iv), we use the RESPACK code (Refs. 44, 45), and we first perform the disentanglement of the medium-energy subspace (the Ni3$d$ and O2$p$ bands near the Fermi level) from the other bands, using the procedure in Ref. 46. Details on the application of this procedure can be found in e.g. Refs 45, 47, 48 in the case of copper oxides. Then, we use the medium-energy subspace as a window to construct the $d_{x^2-y^2}$ and $d_{3z^2-r^2}$ Wannier orbitals. We impose an inner window near the Fermi level so that the Wannier band dispersion matches the GGA band dispersion. The Wannier orbitals are constructed as maximally localized Wannier functions. The bonding character from the lowest bands in the medium-energy subspace is filtered out using the procedure in Ref. 45: Namely, we first construct the Wannier band dispersion for the 11 bands with Ni3$d_{x^2-y^2}$, Ni3$d_{3z^2-r^2}$, in-plane O2$p_\sigma$, and apical O2$p_z$ character, then we discard the dispersion for the 7 lowest bands, keeping only the dispersion for the 4 highest bands. The principal Hamiltonian parameter that is discussed here and shown in Extended Data Fig. 8c is $\Delta E$ (the difference in onsite energies of $d_{x^2-y^2}$ and $d_{3z^2-r^2}$ orbitals). The Fermi surfaces in Extended Data Fig. 8 are calculated using FermiSurfer (Ref. 49).

## Note

During the preparation of this manuscript, we became aware of reports of high-field[50] and high-pressure[51] studies of bilayer nickelate thin films. Ambient-pressure two-step transitions in bilayer nickelate films have been discussed in terms of spin-glass superconductivity[52] and granular superconductivity associated with Josephson-coupled superconducting regions[53].

## Data availability

The data that support the findings of this study are available from the corresponding author upon request.

### Methods-only references (including Note)

38. Momma K. & Izumi, F. Vesta 3 for three-dimensional visualization of crystal, volumetric and morphology data. *J. Appl. Crystallogr.* **44**, 1272–1276, (2011).
39. Giannozzi, P. *et al.* QUANTUM ESPRESSO: a modular and open-source software project for quantum simulations of materials, *Journal of Physics: Condensed Matter* **21**, 395502 (19pp) (2009).
40. Giannozzi, P. *et al.* Advanced capabilities for materials modelling with QUANTUM ESPRESSO, *Journal of Physics: Condensed Matter* **29**, 465901 (2017).
41. Perdew, J. P., Burke, K. and Ernzerhof, M., Generalized Gradient Approximation Made Simple, *Phys. Rev. Lett.* **77**, 3865 (1996).
42. Hamann, D. R. Optimized norm-conserving Vanderbilt pseudopotentials, *Phys. Rev. B* **88**, 085117 (2013).

43. Schlipf, M. & Gygi, F. Optimization algorithm for the generation of ONCV pseudopotentials, *Comput. Phys. Commun.* **196**, 36 (2015).
44. Nakamura, K. *et al*. RESPACK: An ab initio tool for derivation of effective low-energy model of material, *Comput. Phys. Commun.* **261**, 107781 (2021).
45. Morée, J.-B., Hirayama, M., Schmid, M. T., Yamaji, Y. & Imada, M. Ab initio low-energy effective Hamiltonians for the high-temperature superconducting cuprates $Bi_2Sr_2CuO_6$, $Bi_2Sr_2CaCu_2O_8$, $HgBa_2CuO_4$, and $CaCuO_2$, *Phys. Rev. B* **106**, 235150 (2022).
46. Miyake, T., Aryasetiawan, F. & Imada, M., *Ab initio* procedure for constructing effective models of correlated materials with entangled band structure, *Phys. Rev. B* **80**, 155134 (2009).
47. Morée, J.-B. & Arita, R. Universal chemical formula dependence of *ab initio* low-energy effective Hamiltonian in single-layer carrier-doped cuprate superconductors: Study using a hierarchical dependence extraction algorithm, *Phys. Rev. B* **110**, 014502 (2024).
48. Morée, J.-B., Yamaji, Y. & Imada, M. Dome structure in pressure dependence of superconducting transition temperature for $HgBa_2Ca_2Cu_3O_8$: Studies by *ab initio* low-energy effective Hamiltonian, *Phys. Rev. Research* **6**, 023163 (2024).
49. Kawamura, M. FermiSurfer: Fermi-surface viewer providing multiple representation schemes. *Comp. Phys. Commun* **239**, 197 (2019).
50. Hsu, Y.-T. *et al.* Fermi-liquid transport beyond the upper critical field in superconducting $La_2PrNi_2O_7$ thin films, *Nat. Commun.* **17**, 3760 (2026).
51. Li, Q. *et al.* Enhanced superconductivity in the compressively strained bilayer nickelate thin films by pressure. *Nat. Commun.* **17**, 3276 (2026).
52. Ji, H. *et al.* Time-reversal symmetry breaking superconductivity with electronic glass in nickelate (La, Pr, $Sm)_3Ni_2O_7$ films. Preprint at arxiv.org/abs/2508.16412 (2025).
53. Han, Z., Xiang, L., Zhou, X. J., Zhu, Z. Granular Superconductivity in $La_2PrNi_2O_{7-\delta}$. Preprint at arxiv.org/abs/2604.07807 (2026).

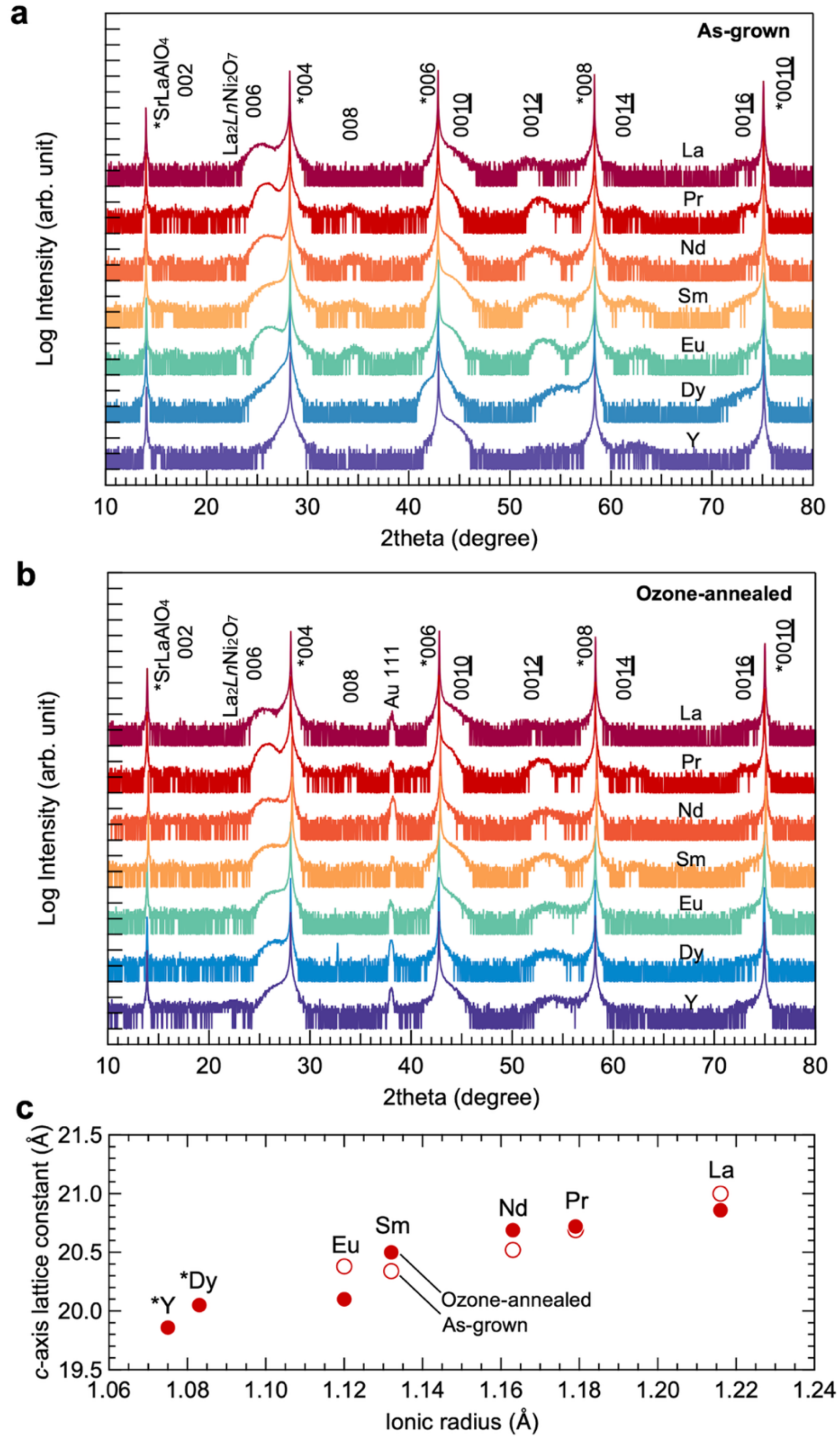

**Extended Data Fig. 1 | X-ray Diffraction of $La_2LnNi_2O_7$ thin films. a,** As-grown $La_2LnNi_2O_7$ films on $SrLaAlO_4$ substrate**. b,** Ozone-annealed $La_2LnNi_2O_7$ films on $SrLaAlO_4$ substrate. *Ln* = La, Pr, Nd, Sm, Eu, Dy, and Y. Au 111-peaks correspond to deposited Au electrodes. **c**, The *c*-axis lattice constant as a function of $Ln^{3+}$ ionic radius[23]. Open and filled circles indicate as-grown and ozone-annealed films, respectively. The *c*-axis values for as-grown and ozone-annealed films were determined from the 006 peaks of $La_2LnNi_2O_7$ using Bragg's law. For the as-grown samples with *Ln* = Dy and Y, the 006 peaks were broad and weak and largely overlapped with the 004 peaks of $SrLaAlO_4$, preventing reliable determination of the *c*-axis lattice parameter.

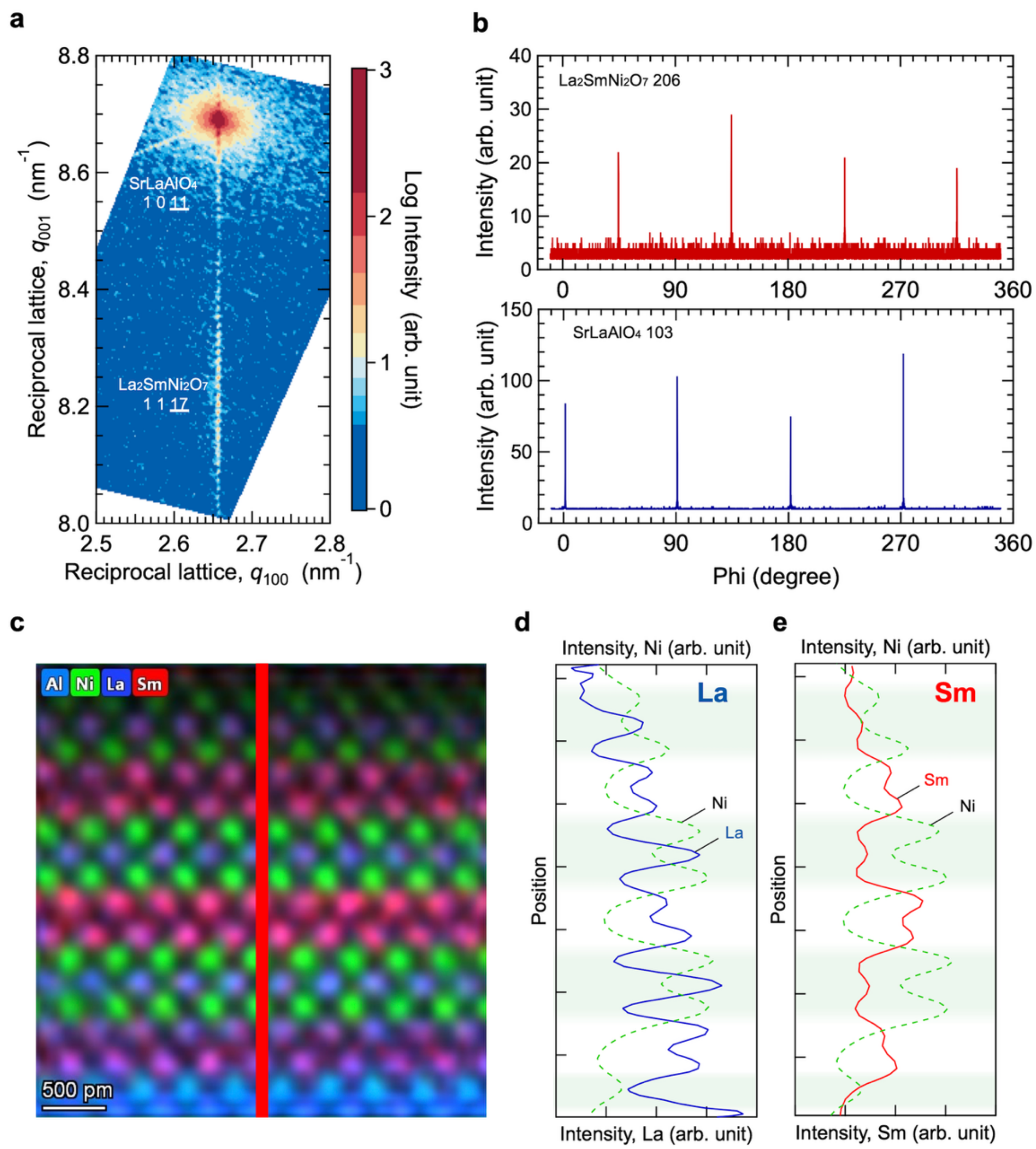

**Extended Data Fig. 2 | Reciprocal space mapping (RSM), phi scans, and Scanning transmission electron microscopy (STEM) of superconducting $La_2SmNi_2O_7$ thin films. a,** RSM of $La_2SmNi_2O_7$ films on $SrLaAlO_4$ substrate**. b,** Phi scans of $La_2SmNi_2O_7$ film and $SrLaAlO_4$ substrate. **c,** High-angle annular darkfield (HAADF) and energy-dispersive X-ray spectroscopy (EDS, for Al, Ni, La, and Sm) images taken along $La_2SmNi_2O_7$ [110] axis. **d** and **e**, Intensity profiles of La and Sm with Ni along the red line in **a**, respectively.

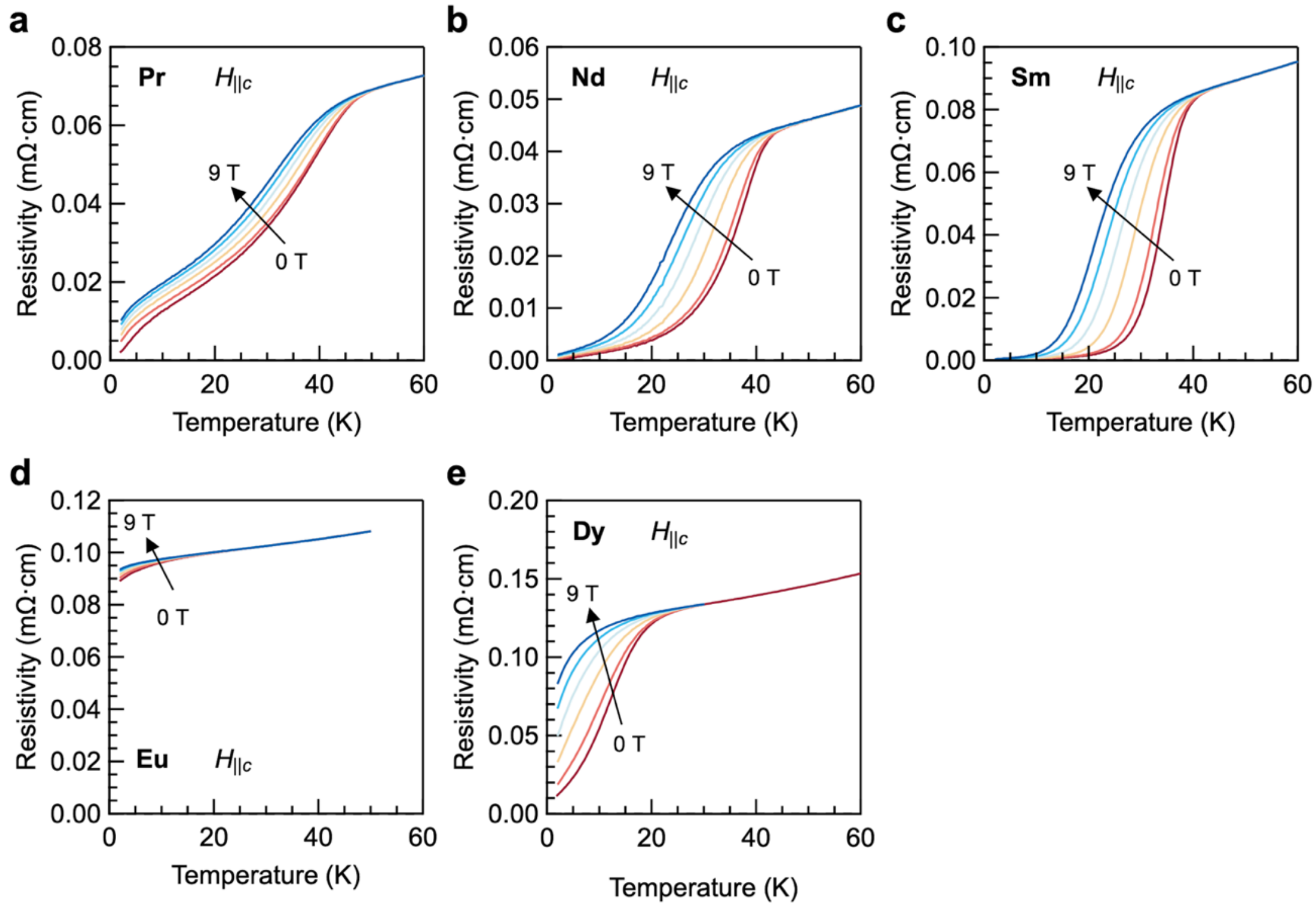

**Extended Data Fig. 3 | Magnetic field responses of $La_2LnNi_2O_7$ thin films. a–e,** Temperature dependent resistivity curves with out-of-plane magnetic fields under 0, 1, 3, 5, 7, and 9 T. *Ln* = Pr, Nd, Sm, Eu, and Dy.

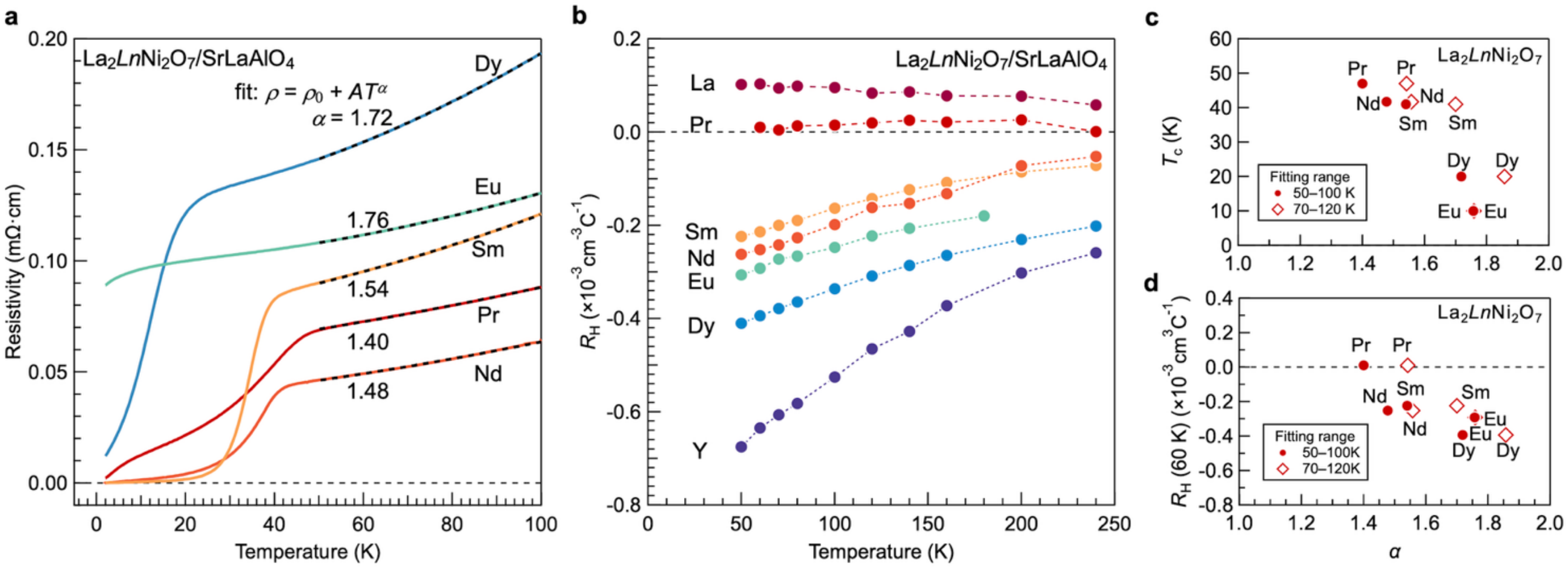

**Extended Data Fig. 4 | Transport properties of $La_2LnNi_2O_7$ thin films. a**, Temperature dependence of resistivity for the $La_2LnNi_2O_7$ films (*Ln* = Pr, Nd, Sm, Eu, Dy). Broken curves represent fits to $\rho = \rho_o + AT^\alpha$. **b**, Temperature dependence of Hall coefficients for $La_2LnNi_2O_7$ thin films (*Ln* = La, Pr, Nd, Sm, Eu, Dy, and Y). **c**, $T_c$ and **d**, Hall coefficients at 60 K plotted as a function of the normal-state resistivity exponent $\alpha$ for the $La_2LnNi_2O_7$ films (*Ln* = Pr, Nd, Sm, Eu, Dy). The $\alpha$ values evaluated over two fitting ranges are shown for comparison with the $\alpha$ values analyzed under high pressure (Extended Data Fig. 7).

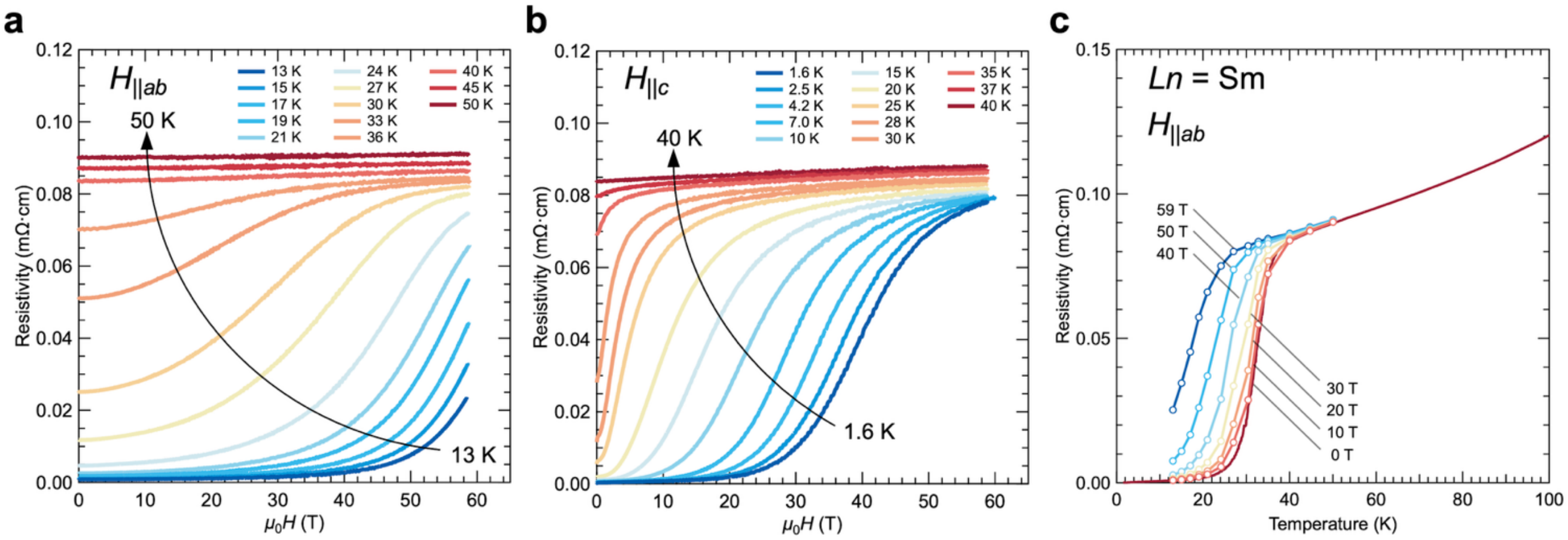

**Extended Data Fig. 5 | High-field electrical transport measurement. a** and **b,** Field dependence of resistivity of $La_2SmNi_2O_7$ film along *ab*-plane and *c*-axis, respectively, in pulsed magnetic fields of up to 60 T. **c**, Temperature dependence of resistivity of $La_2SmNi_2O_7$ film obtained from data shown in **a**.

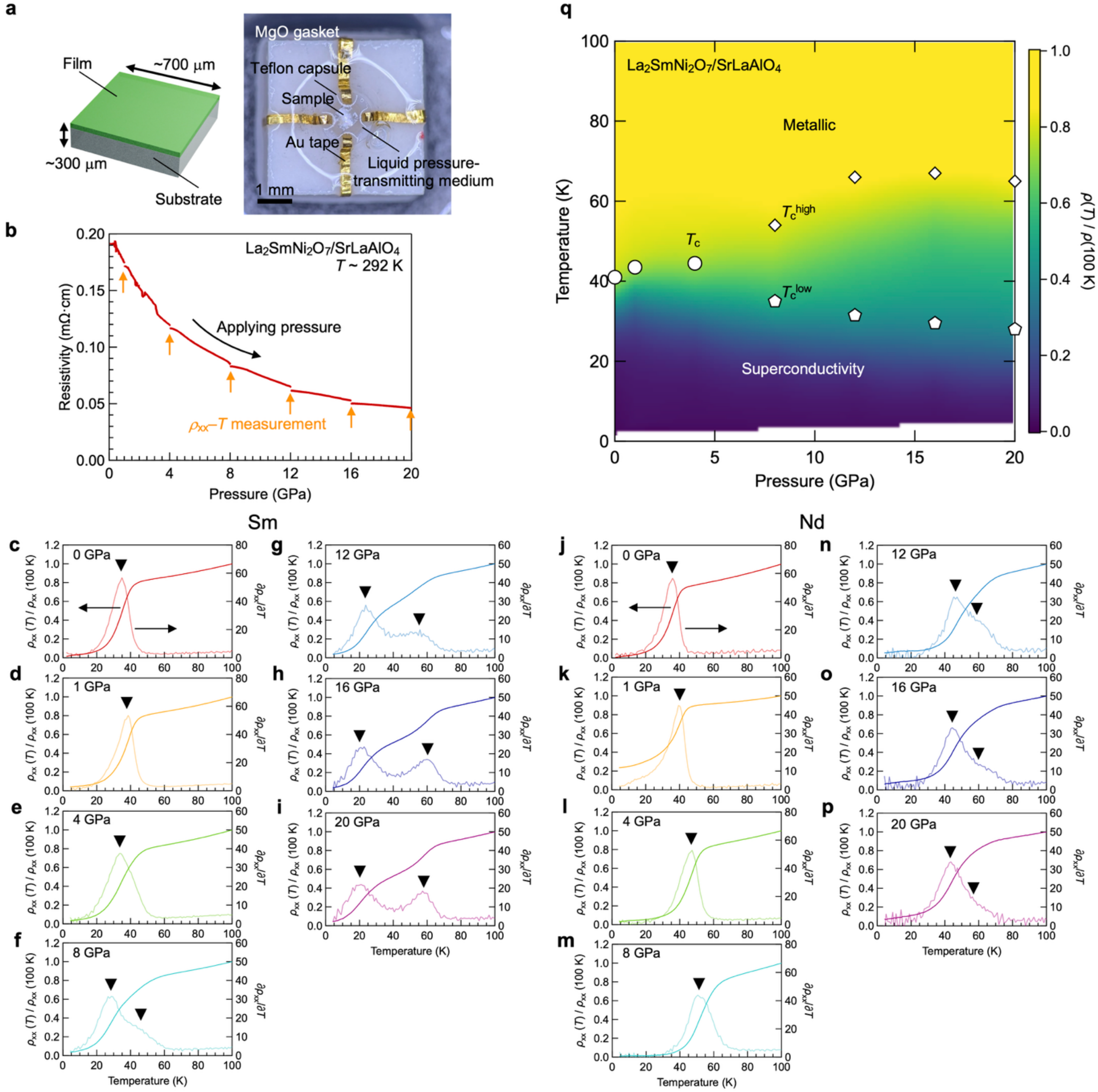

**Extended Data Fig. 6 | High-pressure electric resistivity measurement. a,** Sample geometry for high-pressure electric resistivity measurements (left) and the cubic gasket used for cubic anvil cell type high-pressure measurements (right). **b**, Resistivity near 292 K as a function of applied pressure. Temperature-dependent resistivity was measured at 1, 4, 8, 12, 16, and 20 GPa. The normalized resistivity at 100 K (left axis) and the temperature-dependent deviation of resistivity (right axis) of **c–i,** $La_2SmNi_2O_7$ films and **j–p,** $La_2NdNi_2O_7$ films under 0–20 GPa. Triangles indicate resistive transitions. **q**, A color map of normalized resistivity of a $La_2SmNi_2O_7$ film at 100 K as a function of temperature and pressure.

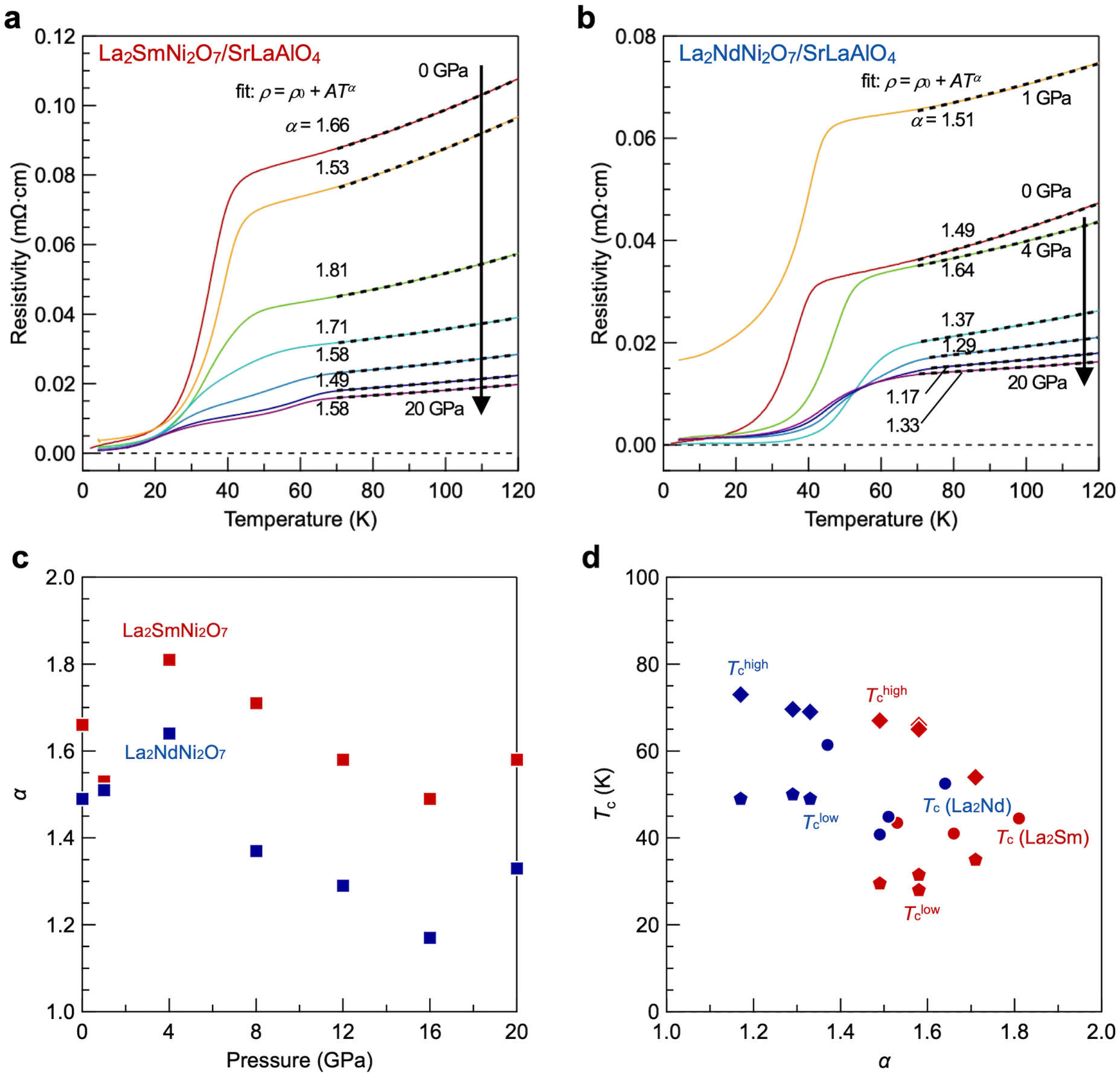

**Extended Data Fig. 7 | Analysis of normal-state transport under pressure.** Temperature dependence of resistivity for **a**, $La_2SmNi_2O_7$ and **b**, $La_2NdNi_2O_7$ films under pressures (0, 1, 4, 8, 12, 16, and 20 GPa). Broken curves represent fits to $\rho = \rho_o + AT^{\alpha}$. The fitting temperature range is 70–120 K, except for $La_2NdNi_2O_7$ films under 12 and 16 GPa (73–120 K), due to the onset $T_c$ of approximately 73 K. **c**, The normal-state resistivity exponent $\alpha$ plotted as a function of pressure for the $La_2SmNi_2O_7$ and $La_2NdNi_2O_7$ films. **d**, The onset $T_c$ (circles) in the low-pressure regime, and $T_c^{high}$ (diamonds) and $T_c^{low}$ (pentagons) in the high-pressure regime, are summarized as a function of $\alpha$.

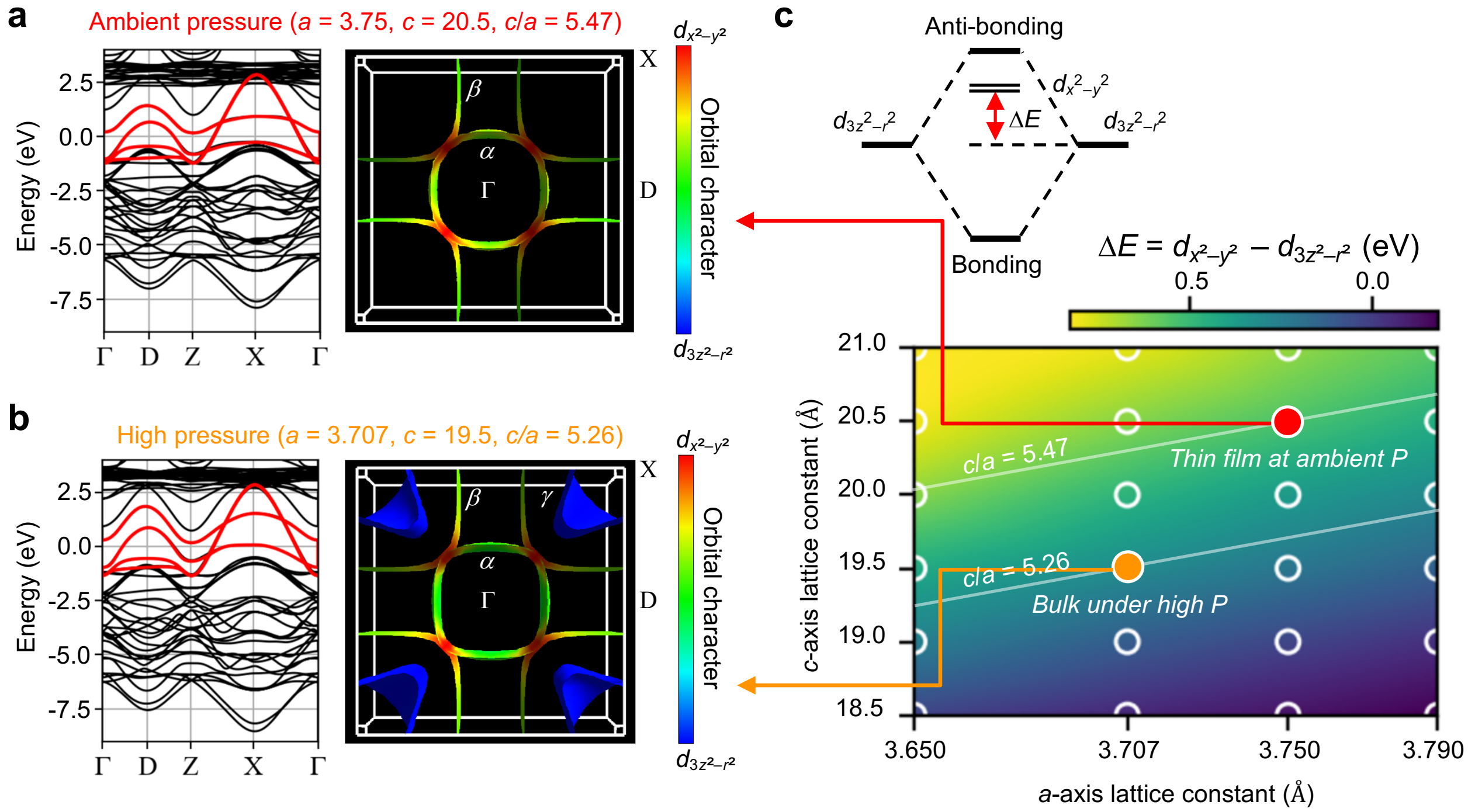

**Extended Data Fig. 8 | Electronic band structure of $La_3Ni_2O_7$ analyzing with different lattice constants. a** and **b,** (left) Calculated electronic band structures for the lattice conditions at typical ambient pressure and at typical high-pressure conditions, respectively. The high-symmetry points in Cartesian coordinates and in units of $2\pi/a$ are: Γ = (0, 0, 0), D = (1/2, 0, 0), Z = (1, 0, 0) and X = (1/2, 1/2, 0). The black and red band dispersions show the GGA and Wannier bands, respectively. The Fermi level is zero. (right) Fermi surfaces corresponding to **a** and **b**, respectively. The color bar shows the momentum-dependent orbital character of the Fermi surface. **c,** Schematic simple electronic structure with $d_{3z^2-r^2}$ and $d_{x^2-y^2}$ and a color plot of $\Delta E$ as functions of $c$-axis and $a$-axis lattice constants. $\Delta E$ is the energy difference between the $d_{x^2-y^2}$ and $d_{3z^2-r^2}$ orbitals in the low-energy Hamiltonian and is correlated to the lattice squeezing. The white open circles correspond to the actual *ab initio* calculations, while other values are obtained by interpolation. Note that the two degenerate $d_{3z^2-r^2}$ orbitals are hybridized through the out-of-plane hopping: Thus, in the band structure, they split into one bonding $d_{3z^2-r^2}$ band and one antibonding $d_{3z^2-r^2}$ band as illustrated here.

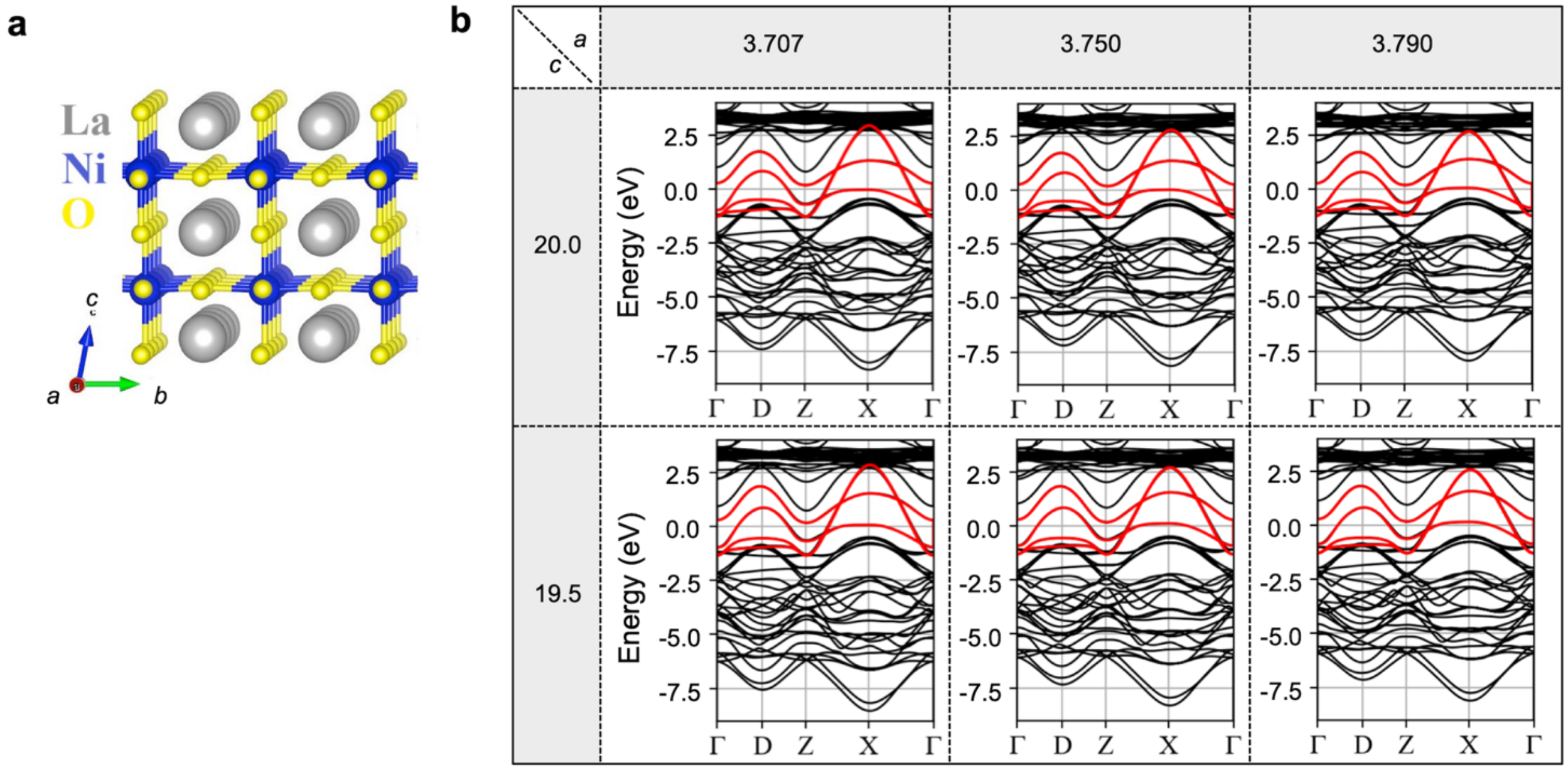

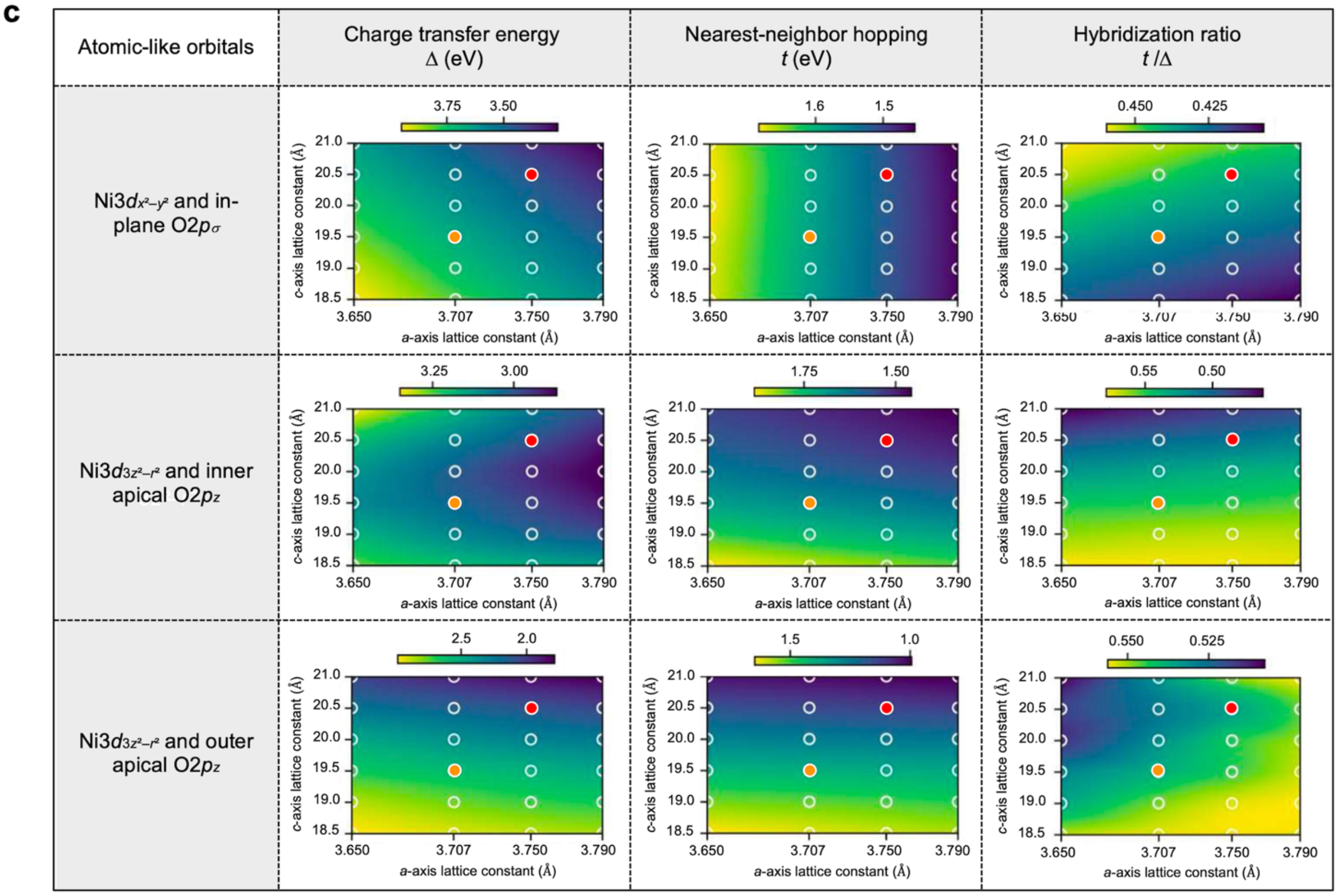

**Extended Data Fig. 9 | Theoretical calculations based on density functional theory. a**, Crystal structure of $La_3Ni_2O_7$. **b**, Dependence of the band structure of $La_3Ni_2O_7$ on the cell parameters *a* and *c* in Å. **c**, Color maps of the charge transfer energy Δ (eV), nearest-neighbor hopping *t* (eV), and hybridization ration *t*/Δ. Closed red and orange circles correspond to ambient-pressure thin-film and high-pressure bulk conditions, respectively. The white open circles correspond to the actual *ab initio* calculations, while other values are obtained by interpolation.

|  | $x$ | $y$ | $z$ |
|---|---|---|---|
| $a$ | $a$ | 0 | 0 |
| $b$ | 0 | $a$ | 0 |
| $c$ | $a/2$ | $a/2$ | $c/2$ |
| **Ni** | 0 | 0 | $\pm d^z_{\mathrm{Ni}}$ |
| **$O_{pl}$ (in-plane O)** | $a/2$ | 0 | $\pm(d^z_{\mathrm{Ni}} + d^z_{\mathrm{O}})$ |
| **$O_{pl}$ (in-plane O)** | 0 | $a/2$ | $\pm(d^z_{\mathrm{Ni}} + d^z_{\mathrm{O}})$ |
| **$La_{in}$ (inner La)** | $a/2$ | $a/2$ | 0 |
| **$La_{out}$ (outer La)** | $a/2$ | $a/2$ | $\pm(d^z_{\mathrm{Ni}} + d^z_{\mathrm{La}})$ |
| **$O_{in}$ (inner apical O)** | 0 | 0 | 0 |
| **$O_{out}$ (outer apical O)** | 0 | 0 | $\pm(d^z_{\mathrm{Ni}} + d^z_{\mathrm{Oap}})$ |

**Extended Data Table 1 | Atom positions for theoretical calculations based on density functional theory.** Primitive vectors *a*, *b*, *c* of Bravais lattice, and atomic positions in the unit cell, in Cartesian coordination.